\documentclass[showpacs, prb,twocolumn,preprintnumbers,superscriptaddress, aps,floatfix]{revtex4-2}

\usepackage{color}
\usepackage{amsmath,amssymb}
\usepackage{pifont}
\usepackage{amssymb}  
\usepackage{bbold}
\usepackage{float}
\usepackage{subfloat}

\usepackage[labelfont=bf]{caption} 
\usepackage[caption=false]{subfig}
\usepackage{tikz}
\usepackage{makecell}
\usepackage{subfig}
\usepackage{pifont}   
\usepackage{graphicx} 
\usepackage{dcolumn}  
\usepackage{bm}       
\usepackage{multirow} 
\usepackage{placeins}
\usepackage[colorlinks]{hyperref}
\usepackage{mathtools}
\usepackage{subcaption}

\usepackage[normalem]{ulem}

\newcommand{\vect}[1]{\mathbf{#1}}
\newcommand{\ket}[1]{\left|{#1}\right\rangle}
\newcommand{\bra}[1]{\left\langle{#1}\right|}

\newcommand{\braket}[2]{\left\langle{#1} | {#2}\right\rangle}

\usepackage{breqn}
\usepackage{mathrsfs}

\begin{document}
\title{Effective K valley Hamiltonian for TMD bilayers under pressure and application to twisted bilayers with pressure-induced topological phase transitions}
\author{Miftah Hadi Syahputra Anfa}
		\affiliation{Physics Department$,$
		King Fahd University
		of Petroleum $\&$ Minerals$,$
		Dhahran 31261$,$ Saudi Arabia}
  \author{Sabri Elatresh}
		\affiliation{Physics Department$,$
		King Fahd University
		of Petroleum $\&$ Minerals$,$
		Dhahran 31261$,$ Saudi Arabia}
      \affiliation{Interdisciplinary Research Center (IRC) for Intelligent Secure Systems$,$ KFUPM$,$ Dhahran$,$ Saudi Arabia}
    \author{Hocine Bahlouli}
		\affiliation{Physics Department$,$
		King Fahd University
		of Petroleum $\&$ Minerals$,$
		Dhahran 31261$,$ Saudi Arabia}
      \affiliation{Interdisciplinary Research Center (IRC) for Advanced Materials$,$ KFUPM$,$ Dhahran$,$ Saudi Arabia}
\author{Michael Vogl}
		\affiliation{Physics Department$,$
		King Fahd University
		of Petroleum $\&$ Minerals$,$
		Dhahran 31261$,$ Saudi Arabia}
      \affiliation{Interdisciplinary Research Center (IRC) for Intelligent Secure Systems$,$ KFUPM$,$ Dhahran$,$ Saudi Arabia}
\begin{abstract}
  
Motivated by recent studies on topologically non-trivial moir\'{e} bands in twisted bilayer transition metal dichalcogenides (TMDs), we study MoTe\textsubscript{2} bilayer systems subject to pressure, which is applied perpendicular to the material surface. We start our investigation by first considering an untwisted bilayer system with an arbitrary relative shift between layers; a symmetry analysis for this case permits us to obtain a simplified effective low-energy Hamiltonian valid near the important $\mathbf{K}$ valley region of the Brillouin zone. \textit{Ab initio} density functional theory (DFT) was then employed to obtain relaxed geometric structures for pressures within the range of 0.0 - 3.5 GPa and corresponding band structures. The DFT data were then fitted to the low-energy Hamiltonian to obtain a pressure-dependent Hamiltonian. We then apply our model to a twisted system by treating the twist as a position-dependent shift between layers - here, we assume rigid layers, which is a crucial simplification. In summary, this approach allowed us to obtain the explicit analytical expressions for a Hamiltonian that describes a twisted MoTe\textsubscript{2} bilayer under pressure. Our Hamiltonian then permitted us to study the impact of pressure on the band topology of the twisted system. As a result, we identified many pressure-induced topological phase transitions as indicated by changes in valley Chern numbers. Moreover, we found that pressure could be employed to flatten bands in some of the cases we considered.
\end{abstract}
\maketitle

\section{Introduction}

In the past few decades, 2-dimensional (2D) materials have attracted much attention from researchers and the scientific communities at large. This surge in interest is justified by their unique properties that often differ significantly from their bulk counterparts. Most crucially, some of their properties might be useful for applications in modern electronic devices. Graphene, for example, has an approximately linear energy dispersion in the low-energy region. This observation implies that electrons mimic the behavior of massless relativistic particles, resulting in a high mobility of the electrons and making graphene attractive for electronic applications \cite{Novoselov2005}. Besides, graphene is mechanically robust, and, at the same time, highly elastic \cite{Lee2008}. This property makes graphene a material that might find application in flexible electronic devices \cite{Ahn2014,Akinwande2014}. Additionally, graphene has outstanding light absorption properties \cite{Bonaccorso2010}, which might lead to potential applications in optoelectronics. Another 2D material that shows promise for technological applications is transition metal dichalcogenides (TMDs). TMDs have a chemical structure that is denoted as $MX_2$ with $M$ being a transition metal atom and $X$ being chalcogens (such as S, Se, or Te) \cite{manzeli20172d}. An interesting effect for 2D TMD structures is that bulk TMDs often display an indirect band gap, whereas a direct band gap exists in its monolayer counterpart \cite{Mak2010,Kuc2011}. Another interesting feature of 2D TMD structures is its strong spin-orbit coupling \cite{Zhu2011}, which may lead to potential applications in spintronic devices \cite{Radisavljevic2011,Duan2014,Zeng2012}. 

Even more exotic physics in 2D materials can be obtained by stacking two or more 2D materials with a slight difference in lattice constant or orientation. Due to the lattice mismatch, a new pattern with long periodicity (moir\'{e} pattern) can emerge. This class of materials has been named moir\'{e} materials and has been reported to exhibit extraordinary properties. For instance, flat bands have been predicted in a graphene bilayer when both layers are twisted relative to one another by a so-called magic angle of $\approx 1.1^\circ$ \cite{Bistritzer2011}. This feature was then confirmed experimentally and observed to give rise to strongly correlated electron behavior \cite{cao2018unconventional, cao2018correlated}, which reflects that kinetic energy has only a very minor contribution to the dynamics of electrons in a flat band. Other interesting findings include the discovery of the so-called Hofstadter's butterfly in twisted bilayer graphene \cite{dean2013hofstadter, hunt2013massive} as well as ferromagnetism \cite{sharpe2019emergent}.

Exotic phenomena have also been studied in twisted TMDs (tTMDs). Experimental works, for instance, report the observation of Mott insulating state \cite{li2021continuous,regan2020wigner} and quantum anomalous Hall effect \cite{li2021quantum}. On the theoretical side, twisted MoTe\textsubscript{2} homobilayers have been predicted to exhibit nontrivial topological moir\'{e} bands \cite{wu2019topological}, which indicates the existence of Chern insulators in this material. This prediction was followed by experimental results of several groups \cite{Anderson2023,Cai2023,Zeng_2023,Park2023,Xu2023,Wang2024,PhysRevX.13.031037}, which confirmed signatures of the theoretical predictions in their data. One might also add another TMD layer to obtain so-called trilayer TMDs. Trilayer TMDs give additional room for modification and have been shown to give rise to a plethora of Chern transitions\cite{Hassan2023}. Twisted TMDs also act as an exciting playground for simulating idealized models such as Hubbard \cite{Wu2018Hubbard,Tang2020} and generalized Kane-Mele models \cite{wu2019topological,devakul2021magic}. In some recent work \cite{Qiu2023}, authors managed to construct an interacting model Hamiltonian. In the same work Ref. \cite{Qiu2023}, they considered different twist angles and managed to predict transitions between various phases, including quantum anomalous Hall insulator (QAHI), quantum spin Hall insulator (QSHI), and various antiferromagnetic phases.

Generally, we stress that these rich properties are highly tunable via the twist angle. However, one has to choose a twist angle once the material is synthesized to study moir\'{e} structures in an actual experimental situation. This requirement makes it difficult to tune the angle afterward. One has to prepare new experimental samples for each different twist angle. It would be useful to tune material properties in a similar way even after a twist angle has been chosen by introducing additional knobs that allow similar tuning to the twist angle. Such an approach is expected to make it possible to tune physical properties in situ without synthesizing a new twisted TMD sheet for every different twist angle. Furthermore, it is also expected to give more control over the system's physical properties. 

This additional knob could be chosen in various ways. For instance, it could be circularly polarized light, which has been studied for twisted bilayer graphene \cite{Topp2019,Katz2020,Yantao2020,Vogl2020,Ikeda2020} and TMDs \cite{Vogl2021}. Another additional knob that can be introduced to the moir\'{e} system is pressure. There are tremendous efforts underway, both experimentally \cite{Yankowitz2019,Szentpeteri2021,Zhang2022,Gao2020} and theoretically \cite{Chittari2018,Carr2018,Padhi2019,Lin2020},  to study the effect of pressure on graphene-based moir\'{e} systems to employ it to tune its physical properties. As for TMDs, Ref. \cite{Morales-Duran2023} studied the effect of pressure on the stability of the fractional Chern insulator phase in twisted WSe\textsubscript{2}. We note that the focus of our current work differs substantially from theirs in that we derive explicit pressure-dependent parameters for a simplified model and in that we only study non-interacting topological phases rather than the interesting interacting phase discussed in their work. Other research groups also try to introduce electric fields \cite{Liu2020} and strains \cite{Hou2024,Yan2013,Bi2019,Koegl2023} to further tune the properties of graphene-based moir\'{e} materials.

In this paper, we want to help simplify the study of tTMDs subjected to pressure. We present an explicit pressure-dependent low-energy Hamiltonian for a twisted MoTe\textsubscript{2} homobilayer in the non-interacting limit to achieve this goal. The model was derived from a general symmetry-based guess \cite{Balents2019}. A first-principle density functional theory (DFT) approach \cite{Hohenberg1964,Kohn1965} was employed to obtain band structures under pressure. In this work, we will focus our attention on the bands in a small region around the $\vect{K}$ valley. Of course, this does not capture all the physics since there is for example contributions from the $\Gamma$ point. However, for the energy range of interest, $\vect{K}$ valleys contribute significantly to the density of states and will therefore play an important role in the physics. Here, band structures were obtained via DFT computations at each pressure. The results were then fitted to a low-energy Hamiltonian, which was obtained from a symmetry analysis, which allowed us to obtain explicit expression of pressure-dependent coupling parameters. We also limit the range of pressures in which our model is valid. Further investigation then leads us to the prediction of pressure-induced topological phase transitions as indicated by jumps in valley Chern numbers that correspond to specific energy bands.

The remainder of the paper is structured as follows. In Section \ref{sec:material-without-pressure}, we present the low-energy Hamiltonian of the twisted TMD and explain the symmetry-based methods that were employed to obtain it. We demonstrate the steps to obtain the coupling parameters by fitting them to DFT data and explain the methods to compute valley Chern numbers and energy bands. Then, in Section \ref{sec:pressurized-case}, we show how we use DFT computations to study structures under pressure and obtain their energy bands. Subsequently, we use this data to find the expression of the pressure-dependent coupling parameters of the Hamiltonian. We further investigate the physics of the pressurized twisted TMD by computing valley Chern number of particular energy bands within the pressure range of 0 - 3.5 GPa for twist angle $\theta$ of $1.4^\circ$ and $2.3^\circ$ and observe topological transitions that occur in our pressure range. Lastly, we conclude by summarizing our main findings and discussing the potential usage of our effective Hamiltonian in TMDs subject to additional perturbations in Section \ref{sec:conclusion}.

\section{Material without pressure}
\label{sec:material-without-pressure}
\begin{figure}
    	\centering
    	\includegraphics[width=\linewidth]{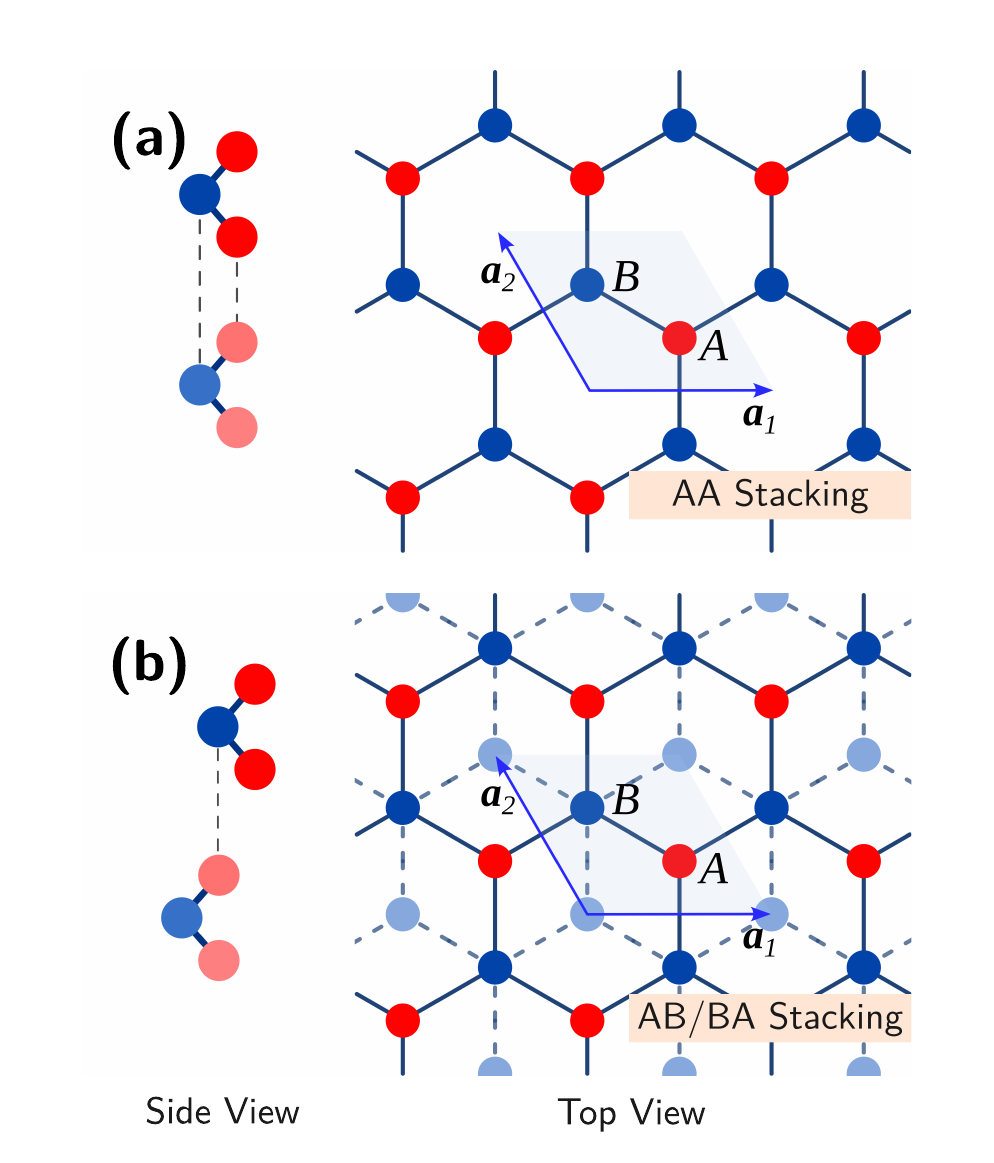}
    	\caption{Illustration of the MoTe\textsubscript{2} homobilayer in (a) AA and (b) AB/BA stacking configurations}
    	\label{fig:bilayer_stacking}
    \end{figure}

This study focuses on MoTe\textsubscript{2} homobilayer systems -including twisted bilayers. Fig. \ref{fig:bilayer_stacking} illustrates such systems for the untwisted case - specifically, the MoTe\textsubscript{2} bilayer in the AA and AB/BA stacking configurations. Before we study the twisted bilayer system, we first discuss the untwisted bilayer case to gain some physical understanding. The discussion in this section is detailed to ensure that the work is both pedagogical and self-contained.

\subsection{Untwisted case and symmetry-based guess for a low energy Hamiltonian for an arbitrary interlayer-shift}
    \label{subsec:no-pressure_untwisted}

\begin{figure*}
	\centering
	\includegraphics[width=0.7\linewidth]{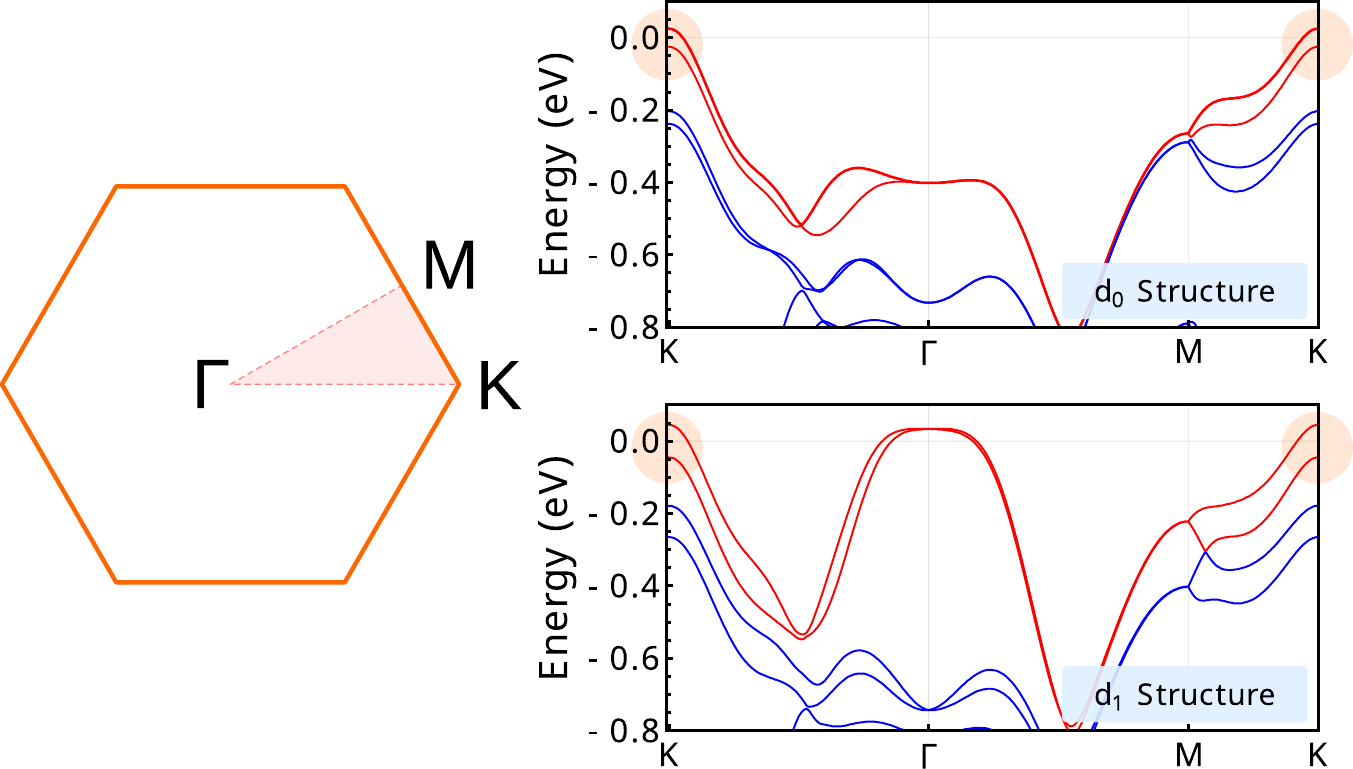}
	\caption{(\textit{Right}) DFT band structures of MoTe\textsubscript{2} bilayer for $\vect{d}_{0}$ displacement and $\vect{d}_{1}$ displacement. The two topmost valence bands (the red lines) close to the $\vect{K}$ valley have a parabolic shape as indicated by a circle we used to mark them. The energy has been shifted to place zero energy between the two topmost valence bands. (\textit{Left}) The Brillouin zone and high-symmetry paths for MoTe\textsubscript{2} homobilayer.}
    \label{fig:dftband}
\end{figure*}

The untwisted bilayer system has a unit cell spanned by lattice vectors 
\begin{equation}
    \label{eq:lattice_constant}
    \vect{a}_1 = a_0 (1,0), \qquad   \vect{a}_2 = a_0 (-1/2, \sqrt{3}/2)
\end{equation}
as illustrated in Fig. \ref{fig:bilayer_stacking}. We note that the unit cell for these systems is smaller than the one of a twisted bilayer - as we will see later. This observation is useful because smaller unit cells make computations less expensive. It also means that we do not have to worry about involving a twist angle as an additional parameter in our discussion yet. These observations make studying the untwisted case a good starting point.

To understand the geometry of layer shifts, we consider the so-called AA stacked MoTe\textsubscript{2} bilayer system first (Fig. \ref{fig:bilayer_stacking}(a) --here, atomic positions of atoms in the $x-y$ plane of one layer are exactly above those in the second layer). We may introduce a displacement vector 
    \begin{equation}
        \label{eq:disp_vec}
        \vect{d}_n = n(-\vect{a}_1 + \vect{a}_2)/3, \quad n=0,\pm 1,
    \end{equation}
which allows the top layer to be shifted with respect to the bottom layer. Choosing $n=\pm 1$ corresponds to the so-called AB/BA stacking (Fig. \ref{fig:bilayer_stacking}(b)) while $n = 0$ corresponds to the AA stacking.

As a first step, we want to obtain the band structure of this system under the three displacements, i.e. $n=0$ and $n=\pm 1$, by performing a density functional theory (DFT) computation. A fully relativistic DFT computation was necessary to obtain accurate band structure results and has been performed using Quantum Espresso \cite{giannozzi2009quantum} with the Local Density Approximation (LDA) exchange-correlation functional and projector augmented-wave (PAW) pseudopotentials. The kinetic energy cutoff for the wave functions and the corresponding charge density were set to 100 Ry and 408 Ry, respectively, where the Brillouin zone was sampled with a $16 \times 16 \times 1$ $\Gamma$-centered Monkhorst-Pack grid of $k$-points  for the step used to obtain self-consistent results.  We also introduced a vacuum layer of thickness $\approx 22$\AA, which allowed us to model TMD bilayers while working with an appropriate DFT code that works with 3D plane waves (at such a distance coupling between a bilayer and spurious copies of the layer is negligible and the computations converge rapidly enough).

We allowed the system to fully relax at a pressure of 0.0 GPa to obtain accurate lattice constants. With this approach, we computed the following lattice constants $a_0$ (see Eq. \eqref{eq:lattice_constant} and Fig. \ref{fig:bilayer_stacking}). For the unshifted case ($\vect{d}_0$) we found $a_0=3.482$ \AA \ while for the shifted case($\vect{d}_{\pm 1}$) $a_0=3.487$ \AA. These numbers are very close to the lattice constants obtained for monolayer MoTe\textsubscript{2} using DFT calculation with the LDA \cite{Liu2013}. They differ only slightly due to coupling between the layers. Additionally, the vertical distance between the Mo atoms of each layer ($D_z$) was found to be 7.79\AA \ for the unshifted structure and 6.91\AA \ for the shifted structures, which also agrees with Ref. \cite{wu2019topological}. We note that we also performed comparisons to GGA\cite{PhysRevLett.77.3865} and GGA with van der Waals corrections\cite{10.1063/1.3382344} but we did not observe very significant changes (similarly for the pressure dependent case later). Hence in what follows we will stick to LDA computations. 
    
The DFT band structures for this material with three displacements (Eq. \eqref{eq:disp_vec} with $n=0$ and $n=\pm 1$) along the $K-\Gamma-M-K$ high-symmetry path are shown in Fig. \ref{fig:dftband}. We note that both $\vect{d}_1$ and $\vect{d}_{-1}$ structures are related by a reflection symmetry along the z-axis, and therefore, their band structures are indistinguishable.

It is clear from Fig. \ref{fig:dftband} that the two topmost energy bands near the so-called $\vect{K}$-point behave approximately quadratically. It is worth noting that, like other TMD materials, MoTe\textsubscript{2} features a strong spin-orbit coupling that causes the splitting of the top valence bands. Particularly, they correspond to spin-up ($\uparrow$) states for the $\mathbf{K}$ valley (spin-down ($\downarrow$) states for $\mathbf{K'}$ valley, which is due to the time-reversal symmetry) \cite{Xiao2012,wu2019topological}. For this reason, we are now interested in finding an effective Hamiltonian that describes these two bands near $\mathbf{K}$ for arbitrary shifts between layers. Note that the model that we are developing in this paper is a simplified one and only focuses on the $\mathbf{K}$ region. As we will discuss later, contributions from other regions of the Brillouin zone - most notably the $\Gamma$ point require a separate treatment.

We switch focus to a derivation of the effective Hamiltonian for the $\mathbf{K}$ region that is based on symmetry considerations. In the cases we consider, we find that one can write a two-band model Hamiltonian as
    \begin{equation}
    	H (\vect{d}_n) = 
    	\begin{pmatrix}
    		-c\vect{k}^2 + \Delta_t		&	\Delta_T \\
    		\Delta_T^{\dagger}			&	-c\vect{k}^2 + \Delta_b
    	\end{pmatrix}
    \end{equation}
which acts on the wave function of the top and bottom layers $(\phi_t, \phi_b)$. (Note: One may confirm that the upper band near $\vect{K}$ has contributions mainly from the top layer, and the lower band has contributions mainly from the bottom layer, which suggests these identifications.)
    
The parabolic behavior of the energy dispersion curve can be described by $-k^2$ while the curve's width can be modified by some number $c$. Such a quadratic curve appears symmetrically in both layers because the corresponding bands of a single layer are quadratic, too. To relate our results to the case of a non-relativistic electron, we set $c = \hbar^2/(2m^*)$ where $m^*$ is interpreted as the effective mass of the electron. We included terms $\Delta_{t,b}$ in the diagonal part of the Hamiltonian, which can be interpreted as on-site potentials for each layer. This term controls the bias between the upper and lower layers. We added $\Delta_T$ in the off-diagonal part to describe inter-layer tunneling. To maintain the hermiticity of the Hamiltonian, one needs to ensure that a Hermitian conjugate relates the off-diagonal terms.

To obtain coupling terms valid for any arbitrary shift between layers, they must satisfy certain symmetry constraints due to the system's geometry. In the following subsections, we will discuss the explicit form of the coupling terms.

\subsubsection{Intra-layer energies and their shift dependence}
To find an expression for the intra-layer energies $\Delta_{t,b}$ for a bilayer with arbitrary shifts, we first need to notice a discrete translational symmetry for the shifts (after certain shifts, the structure repeats). We may, therefore, first consider the discrete translation symmetry of our lattice $\Delta_l (\vect{d} + \vect{R}) = \Delta_l (\vect{d})$ where $\vect{R}$ are the lattice vectors and $l = t,b$ is the index for the layers. As a consequence, they can also be expressed as a Fourier series
    \begin{equation}
    	\label{eq:init_intralayer}
    	\Delta_l(\vect d) = \sum_{\vect{G}_j} V_j e^{i\vect{G}_j \cdot \vect{d}}, \qquad l = t,b.
    \end{equation}

    \begin{figure}
    	\centering
    	\includegraphics[width=0.8\linewidth]{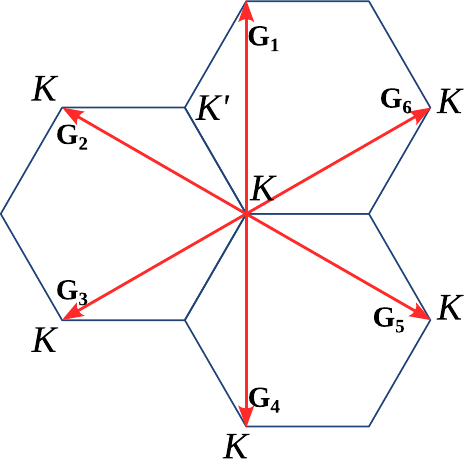}
    	\caption{The reciprocal lattice vectors $\vect{G}_j$ connect point $\vect{K}$ to the nearest equivalent $\vect{K}$ points.}
    	\label{fig:BZ}
    \end{figure}
    
    Here, $\vect{G}_j$ are the reciprocal lattice vectors that connect the origin, which is the $\vect{K}$ valley, to the other equivalent $\vect{K}$ valleys. We only choose the lowest-harmonics $\vect{G}_j$ since, typically, they are the ones that give more contribution to a Fourier series. With this description, $\vect{G}_j$ connects $\vect{K}$ to its equivalent points in the other Brillouin zones. As indicated in Fig. \ref{fig:BZ}, let us choose
    \begin{equation}
        \label{eq:G_vectors}
    	\vect{G}_4 = - \vect{G}_1, \quad \vect{G}_6 = - \vect{G}_3, \quad \vect{G}_2 = - \vect{G}_5,
    \end{equation}
    where the reciprocal lattice vectors $\vect{G}_j$ are defined as
    \begin{equation}
    	\vect{G}_j = \mathsf{R}((j-1)\pi/3)  \frac{4 \pi}{a_0 \sqrt{3}}\hat{j}
    \end{equation}
    and $\mathsf{R}(\theta)$ is a rotation matrix that rotates a vector counterclockwise by an angle $\theta$. 

    The $ C_3 $ symmetry for our lattice structure implies that $ V_j=V $ are the same for all 6 modes (Eq. \eqref{eq:G_vectors}) that we consider. That is, we can then combine Eq.  \eqref{eq:G_vectors} and \eqref {eq:init_intralayer}) to find
    \begin{equation}
    	\label{eq:intralayer}
    	\Delta_l (\vect{d}) = \sum_{j=1,3,5}2V \cos \left( \vect{G}_j \cdot \vect{d} + l\psi \right),
    \end{equation}
    where we have added a phase factor $\psi$ into our function and choose $l = 1$ for the bottom layer and $l = -1$ for the top layer to ensure that there is the possibility for a relative phase shift of $2\psi$ between layers - this does not break rotational symmetry. 

    \subsubsection{Inter-layer Tunneling}
    The next task is to find the expression for the tunneling between layers $\Delta_T$ that fulfills the correct symmetries. Generally, it can be written as
    \begin{equation}
    	\Delta_T(\vect{d}) = \bra{\phi_t} H(\vect{d}) \ket{\phi_b},
     \label{eq:DeltaTasMatrixElement}
    \end{equation}
    where index $t$ ($b$) refers to the top (bottom) layer and $\vect{d} \equiv \vect{d}_t - \vect{d}_b$, which is the difference between the displacement applied to the top and bottom layers. We will now consider a general relative displacement $\vect{d}$.
    
    We again take into account the discrete translational symmetry of our system $\Delta_T (\vect{d} + \vect{R}) = \Delta_T (\vect{d})$, which permits us to write $\Delta_T (\vect{d})$ as a Fourier series,
    \begin{equation}
    	\label{eq:interlayer_fourier}
    	\Delta_T (\vect{d}) = \sum_{\vect{G}_j} \mathsf{T}_{\vect{G}_j} e^{i \vect{G}_j \cdot \vect{d}}.
    \end{equation}
    Another symmetry that we need to account for is the three-fold symmetry for rotations around the center of the hexagon's un-displaced lattice by 120\textdegree (the $C_3$ symmetry). A new position vector $\vect{r}'$  under such a symmetry transformation is obtained by rotating the old position vector $\vect{r}$ according to
    \begin{equation}
    	\vect{r}' = \mathrm{O}_3 \vect{r};\quad \mathrm{O}_3 = \mathsf{R}(120^{\circ}) =
    	\begin{pmatrix}
    		-1/2		&	\sqrt{3}/2 \\
    		-\sqrt{3}/2			&	-1/2
    	\end{pmatrix}.
    \end{equation}
    
   A $C_3$ symmetry operator acts on the displacement vector and the wave function as follows \cite{Balents2019}
    \begin{align}
    	&C_3 \vect{d}_l (\vect{r}) =  \mathrm{O}_3^{-1} \vect{d}_l (\vect{r}') \\
    	&C_3 \phi_l (\vect{r}) = e^{i\vect{G}_2 \cdot \vect{d}_l (\vect{r}')} \phi_l (\vect{r}'),
    \end{align}
    where $l$ refers to the layer and $\vect{G}_2$ is one of the reciprocal lattice vectors.

    Next, we assume that the spatial average of $\Delta_T (\vect{d})$ is non-zero and dominates. This idea means that the constant component of the Fourier series \eqref{eq:interlayer_fourier} is most important. Let us express it as $\mathsf{T}_{\vect{0}} = w$, which is a constant that we will determine later.
    
    Now, we displace the top layer with respect to the bottom layer by $\vect{d}_t (\vect{r}) = \vect{d}$ and leave the bottom layer as is. The invariance under $C_3$ symmetry requires
    \begin{equation}
    		\bra{\phi_t (\vect{r})} H(\vect{d}) \ket{\phi_b (\vect{r})} = \bra{\phi_t (\vect{r})} e^{-i\vect{G}_2 \cdot \vect{d}}  H (\mathrm{O}_3^{-1}\vect{d})  \ket{\phi_b (\vect{r})}.
    \end{equation}
    This component of the Hamiltonian (see Eq. \eqref{eq:DeltaTasMatrixElement}) is given as
    \begin{equation}
    	\Delta_T (\vect{d}) = e^{-i\vect{G}_2 \cdot \vect{d}}  \Delta_T (\mathrm{O}_3^{-1}\vect{d}).
    \end{equation}
    Now, we use the Fourier series \eqref{eq:interlayer_fourier} to obtain
    \begin{equation}
    \label{eq:invariance_G}
    \begin{split}
    	 \sum_{\vect{G}_j'} \mathsf{T}_{\vect{G}_j'} e^{i \vect{G}_j' \cdot \vect{d}} &= e^{-i\vect{G}_2 \cdot \vect{d}}  \sum_{\vect{G}_j} \mathsf{T}_{\vect{G}_j} e^{i \vect{G}_j \cdot \mathrm{O}_3^{-1}\vect{d}} \\
    	 &= \sum_{\vect{G}_j} e^{-i\vect{G}_2 \cdot \vect{d}} \mathsf{T}_{\vect{G}_j} e^{i \mathrm{O}_3 \vect{G}_j \cdot \vect{d}} \\
    	 &= \sum_{\vect{G}_j} \mathsf{T}_{\vect{G}_j} e^{i (\mathrm{O}_3 \vect{G}_j - \vect{G}_2) \cdot \vect{d}}, \\
    \end{split}
    \end{equation}
    from which we can conclude that
    \begin{equation}
    	\label{eq:newG}
    	\vect{G}_j' = \mathrm{O}_3 \vect{G}_j - \vect{G}_2.
    \end{equation}
    We notice that Eq. \eqref{eq:newG} is important because it permits us to identify reciprocal lattice vectors (Fourier modes) that contribute to the Fourier series \eqref{eq:interlayer_fourier} with the same strength. This identification can be done as follows: First, we take some initial $\vect{G}$ for which we know the contribution to the Fourier series. Inserting it into \eqref{eq:newG} then permits us to find a new reciprocal lattice vector $\vect{G}$ that has the same associated Fourier coefficient. The same procedure can then be iterated to find additional $\vect{G}$ until the procedure naturally truncates. Since we already have a coefficient $w$ corresponding to $\vect{G} = (0,0)$ as one of our most important reciprocal lattice vectors, we can use this as the starting point. Using this procedure, we find that there are three reciprocal lattice vectors,
    \begin{gather}
    	\vect{G}_0 = (0,0), \quad 
    	\vect{G}_2 = \frac{4 \pi}{3 a_0} \left( \sqrt{3}/2, 1/2 \right), \\
    	\vect{G}_3 = \frac{4 \pi}{3 a_0} \left( \sqrt{3}/2, -1/2 \right),
    \end{gather}
    which contribute equally with magnitude $w$.
    
    The inter-layer tunneling in Equation \eqref{eq:interlayer_fourier} can now be approximated as
    \begin{equation}
    	\Delta_T (\vect{d}) = \mathsf{T}_{\vect{0}} + \mathsf{T}_{-\vect{G}_2} e^{-i\vect{G}_2 \cdot \vect{d}} + \mathsf{T}_{-\vect{G}_3} e^{-i \vect{G}_3 \cdot \vect{d}}.
    \end{equation}
    We stress that one has to set $\mathsf{T}_{-\vect{G}_j} = \mathsf{T}_{\vect{0}} = w$, such that the invariance condition \eqref{eq:invariance_G} for $\Delta_T (\vect{d})$ can be satisfied. Bringing back our displacement vector \eqref{eq:disp_vec}, we finally have
    \begin{equation}
    	\label{eq:interlayer}
    	\Delta_T (\vect{d}) = w \left(1 + e^{-i\vect{G}_2 \cdot \vect{d}} + e^{-i \vect{G}_3 \cdot \vect{d}}\right).
    \end{equation}

    A complete low-energy Hamiltonian for the general shift, the untwisted case can now be written as
    \begin{equation}
    	\label{eq:aligned_Hamiltonian}
    	H (\vect{d}) = 
    	\begin{pmatrix}
    		-\frac{\hbar^2}{2m^*}\vect{k}^2 + \Delta_t (\vect{d})		&	\Delta_T (\vect{d}) \\
    		\Delta_T^{\dagger} (\vect{d})			&	-\frac{\hbar^2}{2m^*}\vect{k}^2 + \Delta_b (\vect{d})
    	\end{pmatrix},
    \end{equation}
    where the intra-layer potential $\Delta_{t,b}$ and inter-layer tunneling are given by \eqref{eq:intralayer} and \eqref{eq:interlayer}, respectively. We would like to highlight that this result is identical to that of Ref. \cite{wu2019topological}.
    
    It is instructive to see what happens to the Hamiltonian for the high symmetry shifts we discussed earlier. For the unshifted case $\vect{d}_0$, we have $\Delta_t = \Delta_b = 6V \cos(\psi)$ and  we have $\Delta_T = 3w$, which is the maximum strength of interlayer hopping. This value reflects that in an AA stacking configuration, hopping between layers is mostly direct - with atoms located directly on top of one another - and, therefore, the hopping element is at maximum strength. This result is contrasted by the shifted case $\vect{d}_{\pm 1}$, where  $\Delta_T$ becomes zero. The zero value reflects that when layers are not aligned, couplings between them are suppressed.

    Next, we want to find parameters $V, \psi, w,$ and $m^*$ corresponding to a real material. We note that they may be readily obtained by fitting the eigenvalues of Hamiltonian \eqref{eq:aligned_Hamiltonian} for both the unshifted $\vect{d}_{0}$ and shifted $\vect{d}_{\pm 1}$ structures to the energy data from the DFT calculation. Importantly, this fit is performed simultaneously for both configurations using a combined cost function. We stress that we only consider a small region of radius 0.04 $\pi/a_0$ around the $\vect{K}$ valley since we are only interested in the low energy region near it. We want to accurately model the region where the two topmost valence bands behave approximately quadratically. After performing the fit, we found that $(V, \psi, w, m^*) \approx (8.66 \ \text{meV}, -89.97^{\circ}, -8.34\ \text{meV}, 0.63m_e)$ where $m_e$ is the electron mass. Our results differ only slightly from the results in the literature \cite{wu2019topological} because we performed a full relaxation with approximately zero pressure. That is, we obtained in-plane lattice constants while considering the full bilayer systems and did not work with lattice constants obtained for the single-layer system, such as those that could be obtained from Reference \cite{Liu2013}.  While the small corrections to constants are not substantial, they build confidence in literature results and are of value for this reason. We also note that since $V$ is related to the bias between the two topmost valence bands, different choices of DFT flavors can result in different values of $V$ from what is presented here. Indeed, underestimated/overestimated energy gaps are a well-known DFT problem.
    
\subsection{Twisted case}
\label{subsec:twisted-case}

    \begin{figure*}
    	\centering
    	\includegraphics[width=0.8\linewidth]{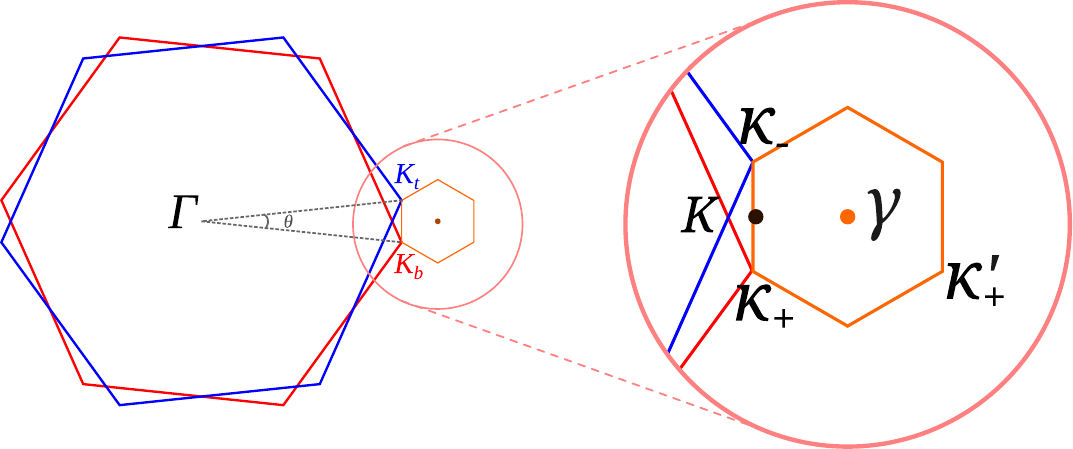}
    	\caption{Illustration of the moir\'{e} Brillouin zone (mBZ), which was constructed by rotating the top and bottom layer Brillouin zone by $+\theta/2$ and $-\theta/2$, respectively, and setting the $\vect{K}$ valley of each BZ as valleys of the newly constructed mBZ named as $\boldsymbol{\kappa}_-$ and $\boldsymbol{\kappa}_+$. The black dot indicates the original position of $\vect{K}$ for the untwisted BZ.}
    	\label{fig:twisted_BZ}
    \end{figure*}
Next, we now want to apply a relative twist between layers. Typically in literature \cite{Balents2019,wu2019topological} this is done by performing a rigid rotation for both layers (or one of them). A more precise treatment would incorporate relaxation effects for the moir\'{e} lattice which has been done for the twisted bilayer graphene \cite{Koshino2020, Vafek2023}. Thus, the model developed in this work offers qualitative insights for physical properties.

Here, we imagine that we can apply the twist by rotating a layer around a Mo atom of the AA-stacked configuration ($d_0$ structure) - we will restrict ourselves to small angles. Such a small twist can be viewed as a collection of local displacements of atoms of the top layer with respect to the atoms of the same type in the bottom layer. Taking only the first order of the Taylor expansion for the rotation matrix, the local displacement $\vect{d}$ can be approximated as $\theta \hat{z} \times \vect{r}$.

    After applying a twist angle, each layer's individual Brillouin zones (BZ) are rotated by $\theta$, as shown in Fig. \ref{fig:twisted_BZ}. We can associate the valley of the top layer $\vect{K}_t$ and bottom layer $\vect{K}_b$ to $\boldsymbol{\kappa}_{\pm}$ points, which are located at corners of a new Brillouin zone, called moir\'{e} Brillouin zone (mBZ). 
    
\begin{figure}
    	\centering
    	\includegraphics[height=4.0cm]{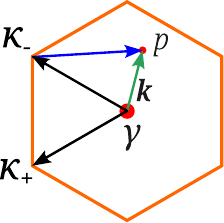}
    	\caption{Illustration of point $p$ in the moir\'{e} Brillouin zone as measured from $\boldsymbol{\gamma}$ and $\boldsymbol{\kappa}_-$.}
    	\label{fig:transform_k}
    \end{figure}

    Due to the twist, a point in momentum space is now measured from two coordinate systems: i.e., $\vect{K}_t$ and $\vect{K}_b$. Measuring it from a common coordinate system will be more convenient; otherwise, it will be difficult to obtain Bloch bands. Here, we set $\gamma$ as the point of origin and set $\kappa_{\pm} = 2\pi \left(-1/\sqrt{3}, \mp 1/3\right)/a_M$ as measured from $\gamma$. As illustrated in Fig. \ref{fig:transform_k}, a momentum $p$ can be expressed as $\vect{k}-\boldsymbol{\kappa}_-$ and $\vect{k}-\boldsymbol{\kappa}_+$. We note in passing that the same prescription can also be obtained if one carefully transforms momentum operators for the local shifts as discussed for the case of twisted bilayer graphene in \cite{Balents2019}.
     
    With these modifications, the moir\'{e} Hamiltonian can be written as
    \begin{equation}
    	\label{eq:moire_Hamiltonian}
    	H (\theta,\vect{r}) = 
    	\begin{pmatrix}
    		-\frac{\hbar^2(\vect{k}-\boldsymbol{\kappa}_-)^2}{2m^*} + \Delta_t (\vect{r})		&	\Delta_T (\vect{r}) \\
    		\Delta_T^{\dagger} (\vect{r})			&	-\frac{\hbar^2 (\vect{k} - \boldsymbol{\kappa}_+)^2}{2m^*} + \Delta_b (\vect{r})
    	\end{pmatrix},
    \end{equation}
    where $\Delta_{t,b}$ and $\Delta_T$ are defined in Eq. \eqref{eq:intralayer} and \eqref{eq:interlayer} by replacing $\vect{d}$ with $\theta\hat{z} \times \vect{r}$, which introduces a position dependence. It is important to note that position dependence is periodic and involves a larger unit cell. This larger unit cell implies a smaller Brillouin zone, which gives rise to moir\'{e} bands, as we will see later. 
    
We should stress that we have made a significant simplification for this section, namely assuming that the lattice structure is rigid under rotation. That is, we neglected the relaxation effects that occur for rotated lattices. In our discussion of the topological phases that follow, we expect that such a simplification will not qualitatively alter our results, given the robustness of topological phases. However, we caution that our model is a simplified one that also does not include interaction effects, and it cannot, therefore, be expected to capture correlated phases such as a fractional Chern insulator phase, like the one discussed in \cite{PhysRevX.13.031037}. 

\subsection{Review of band structure and topology for unpressurized twisted TMD}
\label{subsec:unpressurized-topology}
As a next step before we apply pressure to the twisted TMD, it is useful to review some of its properties to be able to highlight changes. Including those details is useful to keep our work self-contained. The first interesting observable quantity is the band structure. To reveal the band structure of the MoTe\textsubscript{2} moir\'{e} system, we numerically diagonalize the moir\'{e} Hamiltonian \eqref{eq:moire_Hamiltonian} using a plane-wave basis,
    \begin{equation}
        H_{nm} (\vect{k}) = \bra{\phi_n}H(\vect{p + k})\ket{\phi_m}
    \end{equation}
    where $\ket{\phi_n}$ ($\bra{\phi_n}$) is the $n$th plane wave state (plane waves of reciprocal lattice vector momentum have been arbitrarily labeled by integers) and $\vect{k}$ is the crystal momentum ($\vect p$ is the momentum operator). Above, we have applied Bloch's theorem to obtain the momentum space Hamiltonian - particularly the dependence on crystal momentum $\vect{q}$. If we diagonalize the resulting matrix $H(\vect k)$, we find energies $E_n(\vect k)$, i.e., the band structure.

In addition to the band structure, the topological properties of the system can be studied by computing the valley Chern number of several bands. The numerical calculation of the valley Chern number can be performed by following a method proposed by Fukui-Hatsugai-Suzuki \cite{fukui2005chern}. The first step is discretizing the Brillouin zone by $N_x \times N_y$ grid points. In this discretized Brillouin zone, the reciprocal grid vectors are denoted as $\vect{k}_l$ where $l$ refers to the discrete lattice points ($l=1,2,\dots,N_xN_y$). Then, we define the so-called link variable of the $n$-th band,
    \begin{equation}
    	U_{\mu,n} (\vect{k}_l) \equiv \frac{\braket{\phi_n(\vect{k}_l)}{\phi_n(\vect{k}_l + \boldsymbol{\mu}) }}{| \braket{\phi_n(\vect{k}_l)}{\phi_n(\vect{k}_l + \boldsymbol{\mu}) } |},
    \end{equation}
    where $\boldsymbol{\mu} = \hat{\mu}L_{\mu}/N_{\mu}$, $L_{\mu}$ is the length of the lattice and $\hat{\mu}$ a unit vector in the $\mu$ direction.
    
    Then, we define the lattice field strength as
    \begin{equation}
    \begin{split}
    F_{xy} (\vect{k}_l) \equiv &\ln \bigl[ U_{x,n} (\vect{k}_l) U_{y,n}(\vect{k}_l + \boldsymbol{x}) \\
    &\times U_{x,n} (\vect{k}_l + \boldsymbol{y})^{-1} U_{y,n} (\vect{k}_l)^{-1} \bigr].         
    \end{split}
    \end{equation}
    Finally, the valley Chern number of the $n$th band is then calculated as
    \begin{equation}
    	C_n = \frac{1}{2 \pi i}\sum_l F_{xy} (\vect{k}_l).
    \end{equation}
    We note that the valley Chern number as a topological quantity is not just of academic interest, but rather, it is remarkable that it is directly observable through the quantized Hall conductivity of the system as shown by Thouless, Kohmoto, Nightingale, and Nijs (TKNN) \cite{Thouless1982} as
    \begin{equation}
    	\sigma_{xy} = -\frac{e^2}{2 \pi \hbar} \sum_{\alpha \in \text{filled}} C_{\alpha},
    \end{equation}
    where the summation runs over the filled bands.

\begin{figure}[h!]
    	\centering
    	\includegraphics[width=\linewidth]{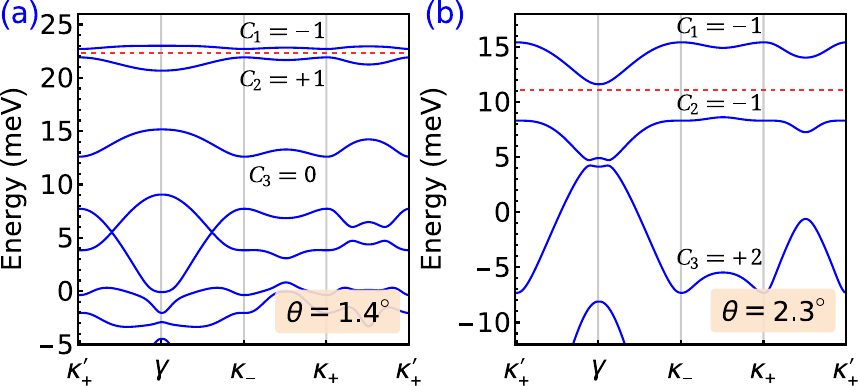}
    	\caption{The moir\'{e} band structure at twist angle (a) 1.2\textdegree \ and (b) 2.3\textdegree. In the figure, we have labeled the valley Chern number of the $n$-th band (counted from the top) as $C_n$.}
    	\label{fig:moireband}
    \end{figure}

 Using the Moir\'{e} Hamiltonian, we can now study the band structure of the twisted TMD system. As an example, we plot the band structure for a twist angle of 1.2\textdegree \ along the $\kappa_+' - \gamma - \kappa_- - \kappa_+ - \kappa_+'$ high-symmetry paths (Fig. \ref{fig:twisted_BZ}) as can be seen in Fig. \ref{fig:moireband}(a). The calculated valley Chern numbers for the two topmost bands are $C_1 = -1$ and $C_2 = 1$ for the first and second band, respectively, which confirms that the valley Chern numbers for both bands are non-trivial. This observation means that the twisted TMD bilayer is a topological insulator when the Fermi level (the dashed line in Fig. \ref{fig:moireband}) lies between the first and the second moir\'{e} bands.
    
\begin{figure}[h!]
    	\centering
    	\includegraphics[width=\linewidth]{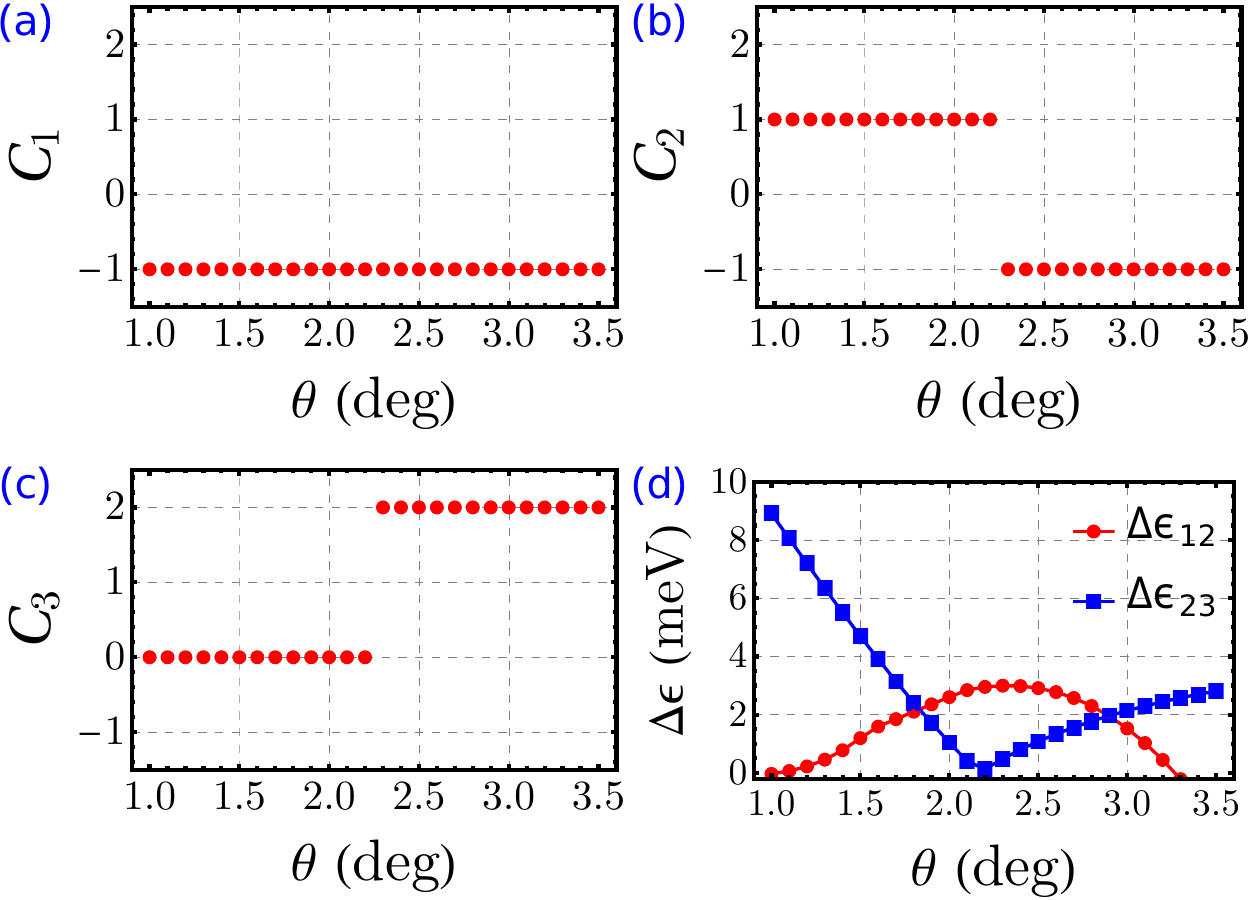}
    	\caption{(a-c) The valley Chern number of the top three bands along various twist angles. For all considered twist angles, the valley Chern number of the first band remains -1, while the second and third bands change valley Chern number around 2.3\textdegree. (d) The energy gap $\Delta \epsilon_{ij}$ between the $i$-th and $j$-th bands is computed as the difference between the minimum energy of the $i$-th band and the maximum energy of the $j$-th band. At around 2.3\textdegree, the gap between the second and third band closes, which is accompanied by a change in valley Chern number.}
    	\label{fig:Chern_numbers}
    \end{figure}

    To better understand how valley Chern numbers change with twist angles, we have plotted the valley Chern number for the first three bands as a function of twist angle $\theta$ in Fig. \ref{fig:Chern_numbers}. Additionally, Fig. \ref{fig:Chern_numbers}(d) shows the gap $\Delta \epsilon_{ij}$ between bands $i$ and $j$ (the difference between the minimum energy of the $i$-th band and the maximum energy of the lower adjacent $j$-th band). From this plot, we can see that the valley Chern number of the first band remains at a value of -1 as we increase the twist angle, while the second and third bands touch and change the valley Chern number from (+1,0) to (-1,+2) at a twist angle of around 2.3\textdegree, where bands 2 and 3 touch. The band structure at the twist angle $\theta = 2.3^\circ$ is shown in Fig. \ref{fig:moireband}(b).
    
    We want to highlight that the results we obtained in this section agree with the existing literature \cite{wu2019topological} with only very slight differences due to the choice of lattice constant that we obtained from a full relaxation of the untwisted structure.
    
\section{Pressurized case}
\label{sec:pressurized-case}

We want to apply pressure to the twisted MoTe\textsubscript{2} bilayer system. As in the case of the system without pressure, we begin our observation from the untwisted system. Our goal is to obtain the coupling parameters for the pressurized systems and use them for the twisted cases by applying a small rotation angle like in the moir\'{e} Hamiltonian Eq. \eqref{eq:moire_Hamiltonian}.

\subsection{Untwisted case}

\begin{figure}[ht]
	\centering
    \includegraphics[width=\linewidth]{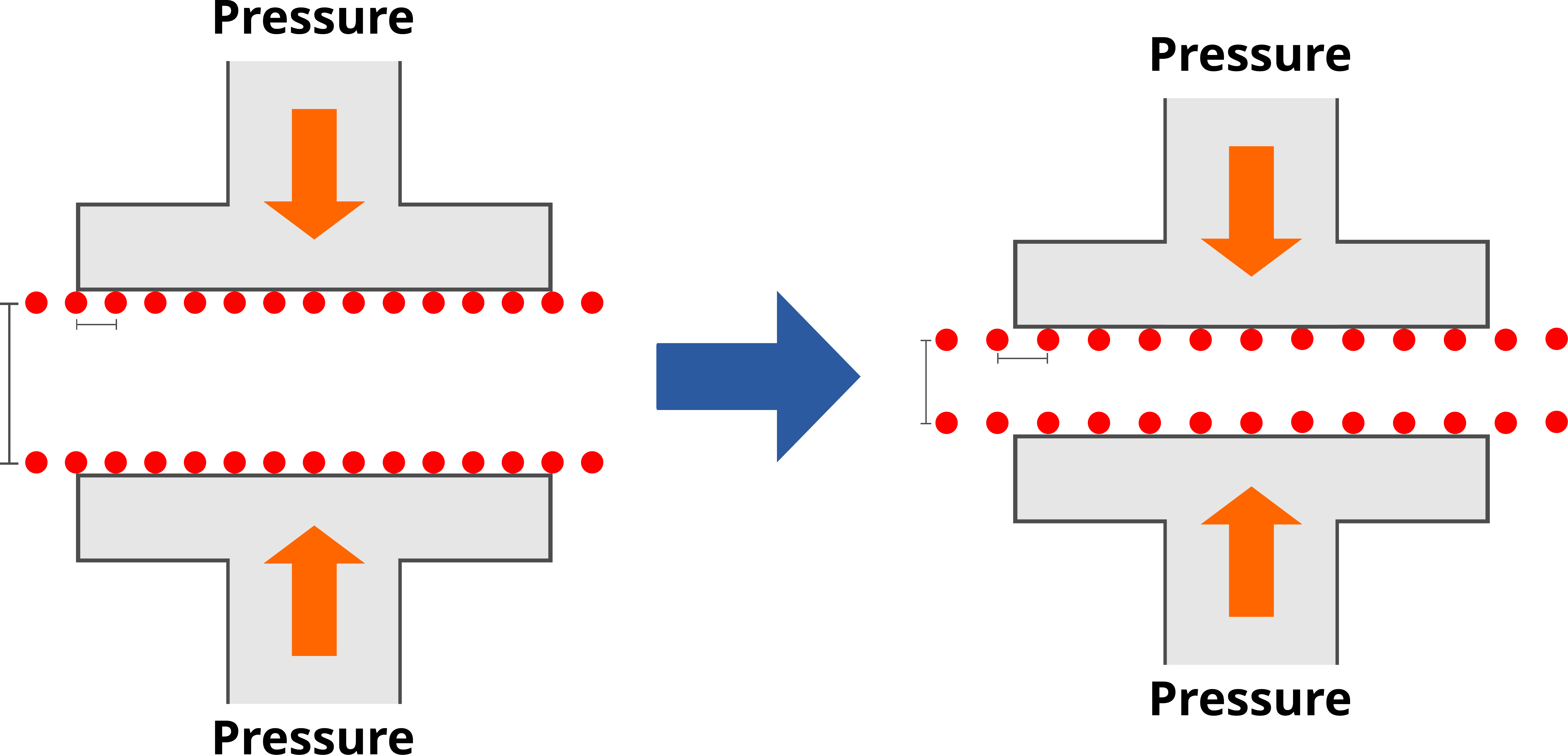}
    \caption{A schematic illustration of a system when it begins to undergo vertical pressure (left) and relaxes after the pressure is applied (right). Intuitively, the applied pressure will push the layers toward each other, reducing the vertical distance between both layers. As atoms from one layer become closer to the other layer, they cause a larger in-plane repulsion force experienced by atoms in the adjacent layer, leading to a larger in-plane lattice constant.}
    \label{fig:pressure_illustration}
\end{figure}

We restrict our study to pressure due to a force applied perpendicular to the bilayer system --it enters as $zz$-component of a stress tensor $\tau_{ij}$. Fig. \ref{fig:pressure_illustration} illustrates this pressure. One might already guess that pressure will lead to structural rearrangements of the system. Intuitively, pressure applied in this way will make the layers closer to each other, leading to a smaller vertical interlayer distance. Moreover, atoms from an adjacent layer will now experience increased in-plane repulsion forces, which, as we expect, will increase the in-plane lattice constant.

One can model this kind of pressure in DFT using the Abinit software \cite{Gonze2020} by setting the value of the $z$-component of the target stress tensor $\tau_{zz}$ in the structural optimization computation - other components have their target set to zero. We note that pressure from the stress tensor can then be calculated as 
\begin{equation}
    P = \frac{1}{3}\sum_i \tau_{ii}    
\end{equation}
(for $i=x,y,z$). 

To obtain pressure-dependent coupling constants for our effective Hamiltonian, band structures for a range of pressures were needed. For each band structure we computed, we relaxed both the unshifted and shifted structures under pressures ranging from 0 - 4 GPa. Our relaxations were based on the Broyden–Fletcher–Goldfarb–Shanno (BFGS) algorithm \cite{BROYDEN1970, Fletcher1970, Goldfarb1970, Shanno1970}. The computations were performed under the fully relativistic scheme using PAW pseudopotentials \cite{Torrent2008} and Perdew-Wang LDA exchange-correlation functional \cite{Perdew1992, Marques2012}. The wave functions energy cutoff was set to 24 Ry, which ensured that the self-consistent field (SCF) step converged.  Moreover, we chose a $\Gamma$-centered $16 \times 16 \times 1$ grid of $k$-points for the SCF calculations. The tolerance on the potential residual was set to $2 \times 10^{-12}$ Ry as the convergence criterion for the SCF calculation, whereas the convergence for the relaxation was based on the maximum absolute force on the atoms, which was set to $2.0 \times 10^{-4}$ Ry/\AA.

\begin{figure}
    \centering
    \includegraphics[width=\linewidth]{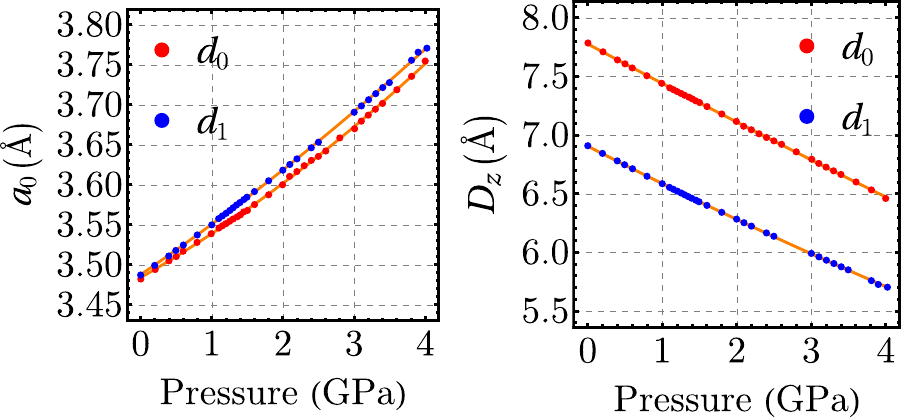}
    \caption{Plots of lattice constants $a_0$ and vertical layer distances $D_z$ as function of pressure. The dots represent the computed data, and the orange lines represent fit functions, i.e., Eq. \eqref{eq:a0_d0} -- \eqref{eq:Dz_d1}}.
    \label{fig:structure_press}
\end{figure}

\begin{figure*}
    \centering
    \includegraphics[width=\linewidth]{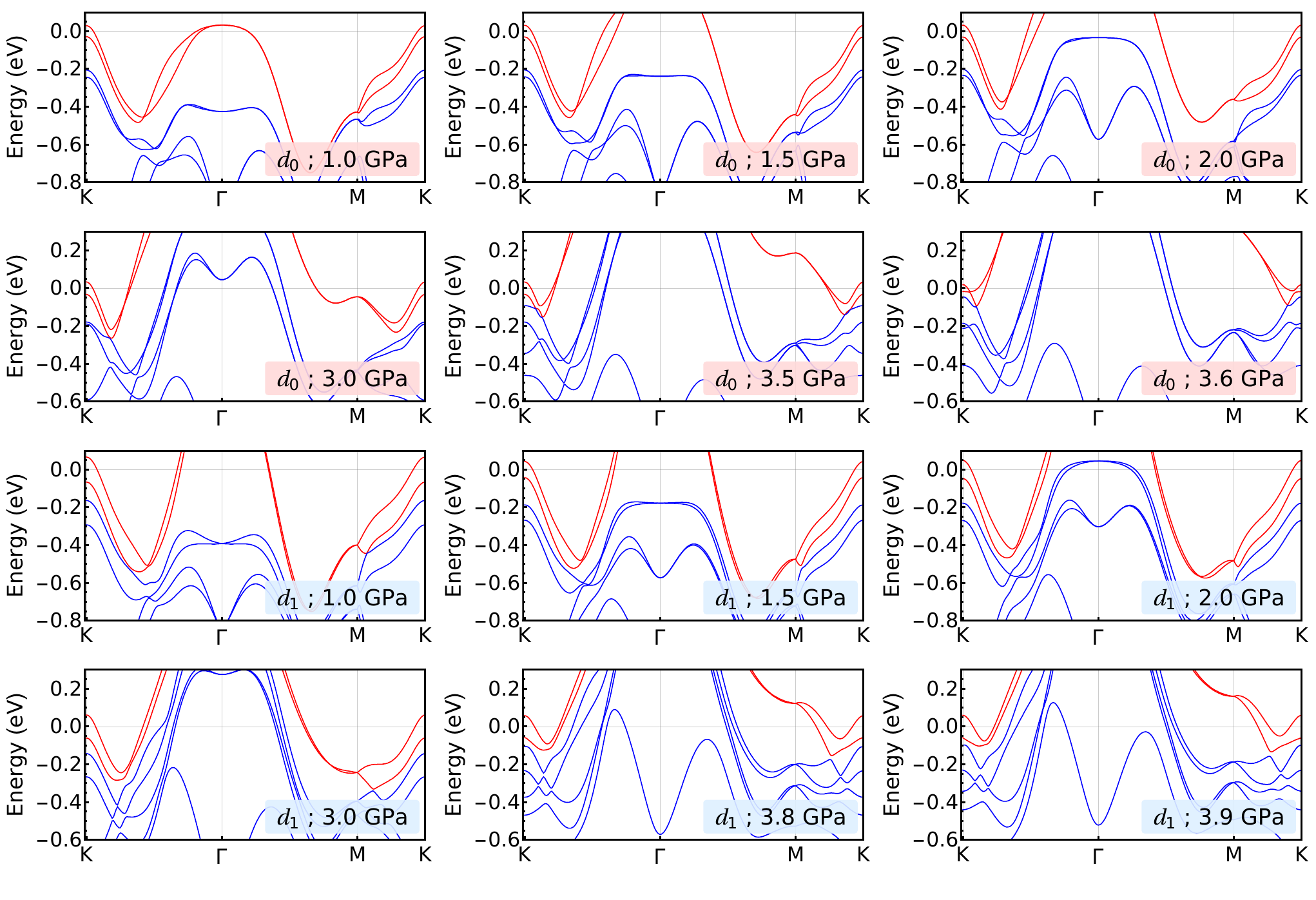}
    \caption{DFT band structure of $d_0$ and $d_1$ structures at different pressures. The two topmost valence bands (indicated by the red line) still behave as quadratic curves around the $\vect{K}$ valley until it breaks down at 3.6 GPa for $d_0$ and 3.9 GPa for $d_1$.}
    \label{fig:dft_bands}
\end{figure*}

Fig. \ref{fig:structure_press} shows the relation between the lattice constant $a_0$ and the vertical distance between layer $D_z$ and pressure. From the data, we can write the lattice constant of $d_0$ and $d_1$, $a_0^{(d_0)}$ and $a_0^{(d_1)}$, as functions of pressure as
\begin{align}
    \label{eq:a0_d0}
    a_0^{(d_0)} (p) &= 3.484 + 0.051 p + 0.004 p^2 \\
    \label{eq:a0_d1}
    a_0^{(d_1)} (p) &= 3.488 + 0.060 p + 0.0025 p^2 ,
\end{align}
respectively. We notice that although both $a_0$ are different, the difference is insignificant (less than 0.4\% for the pressure range of interest). Thus, for the calculation in section \ref{subsec:twisted_press}, we will approximate using $a_0^{(d_1)}(p)$. This also agrees with previous work from the literature in the unpressurized case that also neglected these differences. This step, we stress, is important because it considerably simplifies our analysis.

We also obtain the inter-layer distance of $d_0$ and $d_1$, $D_z^{(d_0)}$ and $D_z^{(d_1)}$, as
\begin{align}
    \label{eq:Dz_d0}
    D_z^{(d_0)} (p) &= 7.78 - 0.34 p + 0.003 p^2 \\
    \label{eq:Dz_d1}
    D_z^{(d_1)} (p) &= 6.91 - 0.33 p + 0.007 p^2 ,
\end{align}
respectively. In the above, we have used a unitless pressure variable $p$ where
$$p = \frac{P}{1 \ \text{GPa}}$$
and $P$ in GPa.

With the relaxed structural parameters, we then compute the band structure in Quantum Espresso using the same simulation parameters as discussed in Sec. \ref{subsec:no-pressure_untwisted}. The DFT band structures of both $d_0$ and $d_1$ systems at various pressures are shown in Fig. \ref{fig:dft_bands}.

A vital idea we checked in our computations was that the honeycomb structure of the lattice for both shifted systems was kept intact during the relaxation (at least for our pressure range). Moreover, we made sure that the two topmost valence bands remained quadratic curves near the $\vect{K}$ valley (this was true up to 3.5 GPa). This property enabled us to adapt our low-energy Hamiltonian Equation \eqref{eq:aligned_Hamiltonian} to structures under pressure. We note that as seen in Fig. \ref{fig:dft_bands}, the quadratic behavior of the topmost valance bands breaks down at the pressure of 3.6 GPa for the $d_0$ structure and 3.9 GPa for the $d_1$ structure. Based on this result, we stress that our model is only reliable for pressure in the range of 0.0 - 3.5 GPa.

Another important observation is that after a certain critical pressure, bands in the $\boldsymbol{\Gamma}$ region enter our energy window of interest. A more detailed analysis of the $\boldsymbol{\Gamma}$ region, would therefore be needed to capture the complete physical picture at such pressure. This observation offers a great opportunity for a follow-up work with exciting physics - for instance, there could be an exciting interplay between $\boldsymbol{\Gamma}$ and $\vect{K}$ valley electrons that is mediated by short-range deformations. However, this is beyond the scope of the current work, where we mostly focus on electronic contributions from the $\mathbf{K}$ point region.

\begin{figure*}
    \centering
    \includegraphics[width=0.8\linewidth]{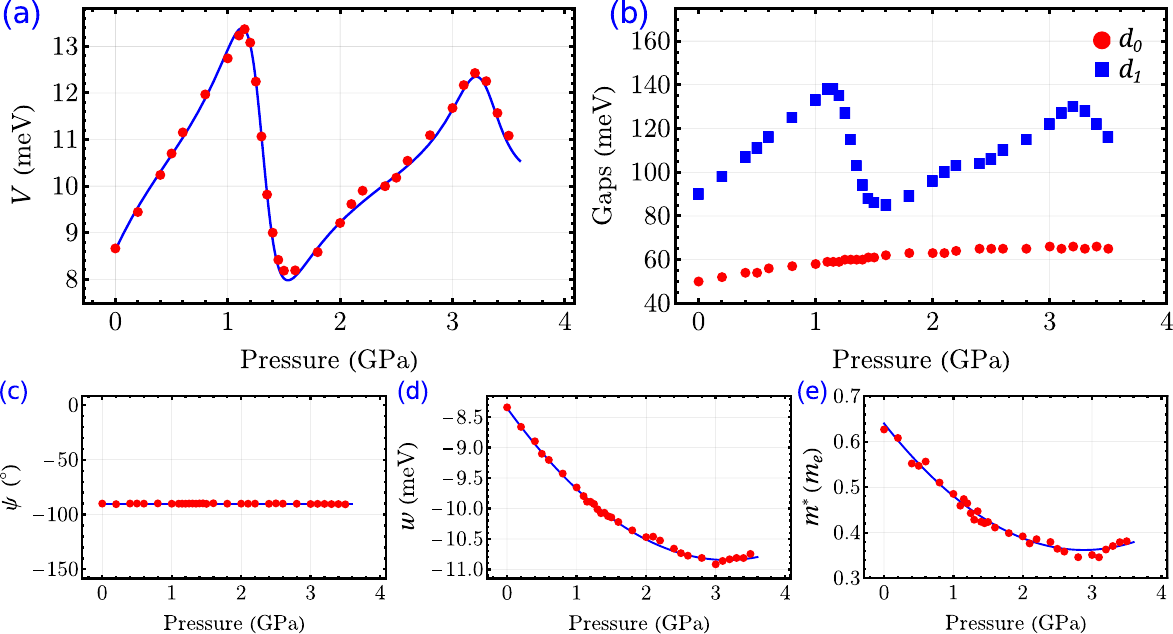}
    \caption{(a, c, d, e) Plots of the coupling parameters as function of pressure. The data is marked as red dots, while the blue line is our fit function. (b) The energy difference between the two topmost valence bands of $d_0$ and $d_1$ structures. Since parameter $V$ corresponds to the gap between the two bands, its oscillation-like behavior originates from the oscillation of the energy gap with pressure variations.}
    \label{fig:param_fit}    
\end{figure*}

As in Sec. \ref{subsec:no-pressure_untwisted}, we can fit the energy data of the region around the $\vect{K}$ valley obtained from the DFT computations to the effective Hamiltonian \eqref{eq:aligned_Hamiltonian} to obtain the coupling parameters at each pressure. The plot of the coupling parameter $V$ as a function of pressure is shown as red dots in Fig. \ref{fig:param_fit}(a). The data could be fitted very well to a function of the form:
\begin{equation}
    \label{eq:V_press}
    V(p) = \left[ \frac{a + b p + c p^2 + d p^3 + e p^4 + f p^5}{g + h p + i p^2 + j p^3 + k p^4 + l p^5} \right] \ \text{meV}.
\end{equation}

with parameters
\begin{equation*}
    \begin{split}
        &a=-694; \quad b=258; \quad c=1327; \quad d= -1405;\\
        &e=486.1; \quad f= -55.8; \quad g=-80.5;\quad h=79.7; \\ 
        &i=61.9;\quad j=-102.3;\quad k=39.51;\quad  l =-4.783;
    \end{split}
\end{equation*}
which were obtained by a least-square fit.

To check our result for additional consistency, we recall that the parameter $V$ corresponds to the gap between the two topmost valence bands. Indeed, this can be seen by comparing the form of $V$ to the energy gap between the two topmost valence bands, as shown in Fig. \ref{fig:param_fit}(b). As expected, both $V$ and the gap have the same pattern.

Fig. \ref{fig:param_fit}(c) shows the plot of $\psi$ at different pressures. Its value appears to be approximately constant along the given pressure. From the fitting, we obtain
\begin{equation}
    \label{eq:psi_press}
    \psi (p) = -90.0797 ^\circ.
\end{equation}

The last two parameters, i.e., $w$ and $m^*$, are shown in Fig. \ref{fig:param_fit}(d--e) and seem to have similar behavior. After performing the fitting, we obtain $w$ and $m^*$ as 
\begin{align}
    \label{w_press}
    &w(p) = \left(-8.35 - 1.58 p + 0.25 p^2 \right) \ \text{meV} \\
    \label{mf_press}
    &m^* (p) = \left(0.64 - 0.193 p + 0.0335 p^2 \right) m_e.
\end{align}
We emphasize that the pressure-dependent couplings given above constitute one of the key results of our paper. Moreover, it is important to note that while the result in the current paper is primarily used to study twisted systems, its application extends beyond this. Indeed, the approach can be used, for instance, to model position-dependent pressure configurations—such as those obtained by pressing on the system via a needle. Similarly, one could use the approach to study bilayers with partial dislocations and pressure acting simultaneously. Of course, it is important to ensure that deformations happen on large enough distance scales such that they do not allow electrons to tunnel from the K valley to other valleys. The model, therefore, should be useful in an understanding of mesoscopic physics.
\subsection{Contribution from Beyond K Region}

As already shown in Figure \ref{fig:dft_bands}, bands near the $\boldsymbol{\Gamma}$ point enter the energy range of interest at higher pressure. To better understand contributions to the physics from the $\vect{K}$ valley region we analyze the density of states (DOS) of the system.

\begin{figure}[ht]
	\centering
	\includegraphics[width=0.6\linewidth]{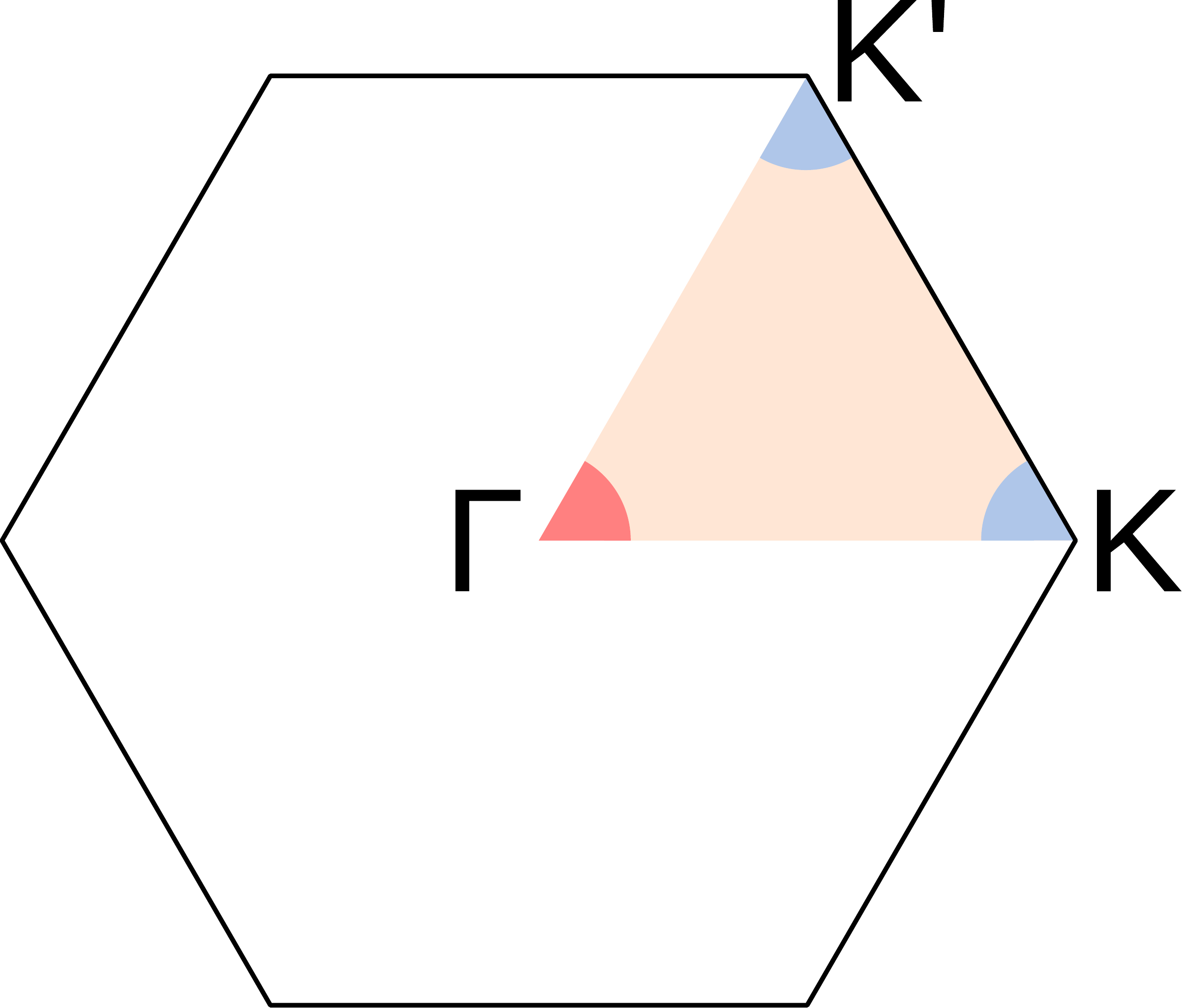}
	\caption{The region used for computation of the density of states. We take 5151 uniformly distributed points in this region for our DOS calculation. Then, we divide them into three subsets: $\vect{K}$ (and $\vect{K'}$) region, $\boldsymbol{\Gamma}$ region, and the outside region.}
	\label{fig:onesegment_BZ}
\end{figure}

Here, after performing an SCF calculation using Quantum Espresso like in Section \ref{sec:pressurized-case}, we compute the energy of the $d_0$ structure at different pressures. As representative examples, we discuss the cases for pressures 0.0 and 1.8 GPa. Due to a sixfold rotational symmetry, we can focus only on one segment of the Brillouin zone, for example, the region enclosed by $\vect{K}-\boldsymbol{\Gamma}-\vect{K'}$ as seen in Fig. \ref{fig:onesegment_BZ}. We compute the energy for 5151 uniformly distributed $k$-points of this region and then divide the energy data into three subsets; the energy that came from the $\vect{K}$ region, the $\boldsymbol{\Gamma}$ region, and outside of both regions. We include all points within the radius of 0.1 (in the unit of $2\pi/a_0$) from $\mathbf{K}$ and $\boldsymbol{\Gamma}$.

The density of states at energy $E$ is then defined as
\begin{equation}
    g(E) =  \sum_{n,\mathbf{k}} \delta(E - E_n (\mathbf{k})),
\end{equation}
where $\delta$ is the Dirac delta function, and $E_n (\mathbf{k})$ is the energy of the $n$-th state at momentum $\mathbf{k}$. 

\begin{figure}[ht]
	\centering
	\includegraphics[width=\linewidth]{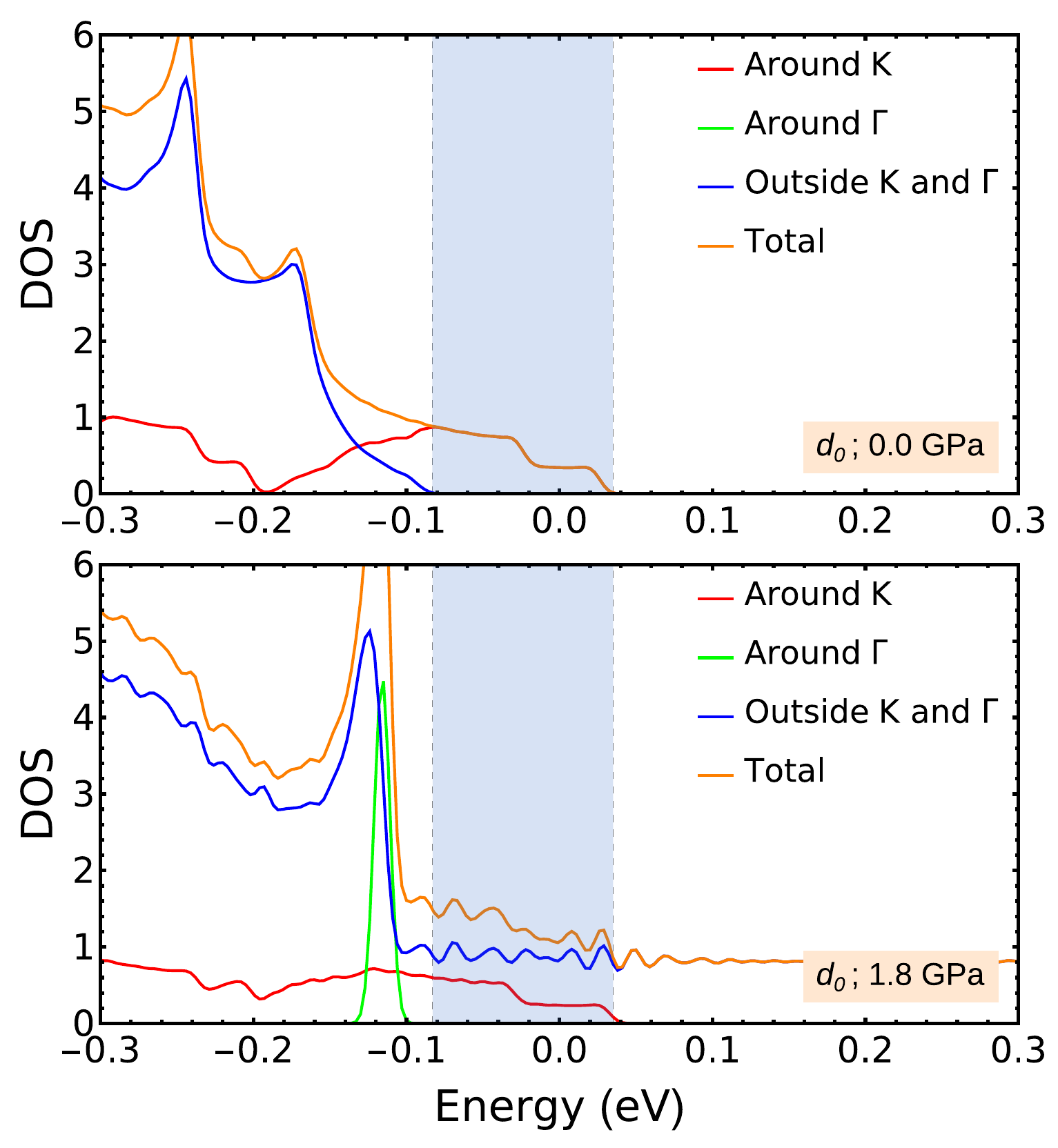}
	\caption{The density of states plot of $d_0$ structure at 0.0 and 1.8 GPa with the region marked in blue is the low-energy region where the quadratic momentum terms in our effective Hamiltonian provide a good approximation. At 0.0 GPa, the $\vect{K}$ has the dominant contribution while at 1.8 GPa it gives about 29.9\% contribution to the total DOS. Also, for both cases, the contribution from the $\boldsymbol{\Gamma}$ region is equal to zero in this energy region.}
	\label{fig:dos}
\end{figure}

To obtain an interpretable plot, it is customary to approximate the Dirac delta function with a Gaussian function,
\begin{equation}
    \delta(E - E_n (\mathbf{k})) \approx \frac{1}{\sqrt{2 \pi}\sigma}\exp{\left( -\frac{(E - E_n (\mathbf{k}))^2}{2\sigma^2} \right)}
\end{equation}
where $\sigma$ is a smearing parameter. The total DOS of the system and the contributions from each region are shown in Figure \ref{fig:dos}.

The plot of the DOS reveals that at zero pressure, the dominant contribution for the DOS in the energy range that we are interested in comes from the region around $\vect{K}$. This means that many physical properties of the system can be expected to have a large contribution from electrons with momentum in this region of the Brillouin zone. However, at higher pressure, the contribution from this region becomes less dominant - although it remains important which amounts to around 29.9\%. We also note that for both cases, the contribution from the $\Gamma$ region is virtually zero in the energy range of our interest. The result for the $d_1$ structure also shows a similar trend. Hence, further analysis of the contribution from other regions is needed to capture the complete physics of the system. Although our simplified model could not capture the full story of the system, it is expected to provide significant insights into parts of the physics - most importantly $\vect{K}$ valley Chern numbers, which can be useful in the valleytronics context \cite{Scuri2020,Belayadi2023}.
    
\subsection{Twisted Case}
\label{subsec:twisted_press}

Since there is no change in the lattice structure due to pressure within the above range, we can still introduce a small twist angle the same way as we did in Sec. \ref{subsec:twisted-case}, i.e., by replacing $\vect{d}$ in Eq. \eqref{eq:intralayer} and \eqref{eq:interlayer} with $\theta \hat{z} \times \vect{r}$. We now obtain the pressure-dependent moir\'{e} Hamiltonian by using the couplings from Eq. \eqref{eq:V_press} - \eqref{mf_press} in Eq. \eqref{eq:moire_Hamiltonian}.

\subsubsection{\texorpdfstring{Twist Angle $\theta = 1.4^\circ$}{Twist Angle 1.4\textdegree}}

\begin{figure}
	\centering
	\includegraphics[width=\linewidth]{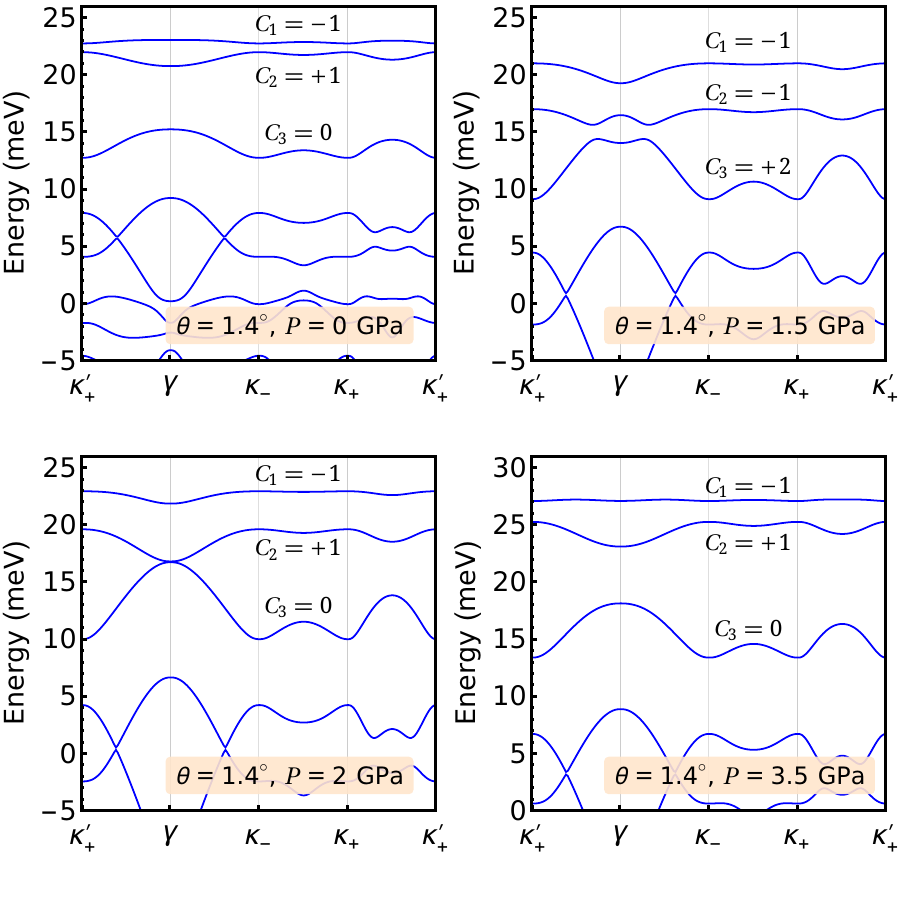}
	\caption{The band structure at various pressures for twist angle $\theta = 1.4^\circ$. In the figure, we have labeled the valley Chern number of the $n$-th band (counted from the top) as $C_n$.}
	\label{fig:bands_1.4deg}
\end{figure}

\begin{figure}
	\centering
	\includegraphics[width=\linewidth]{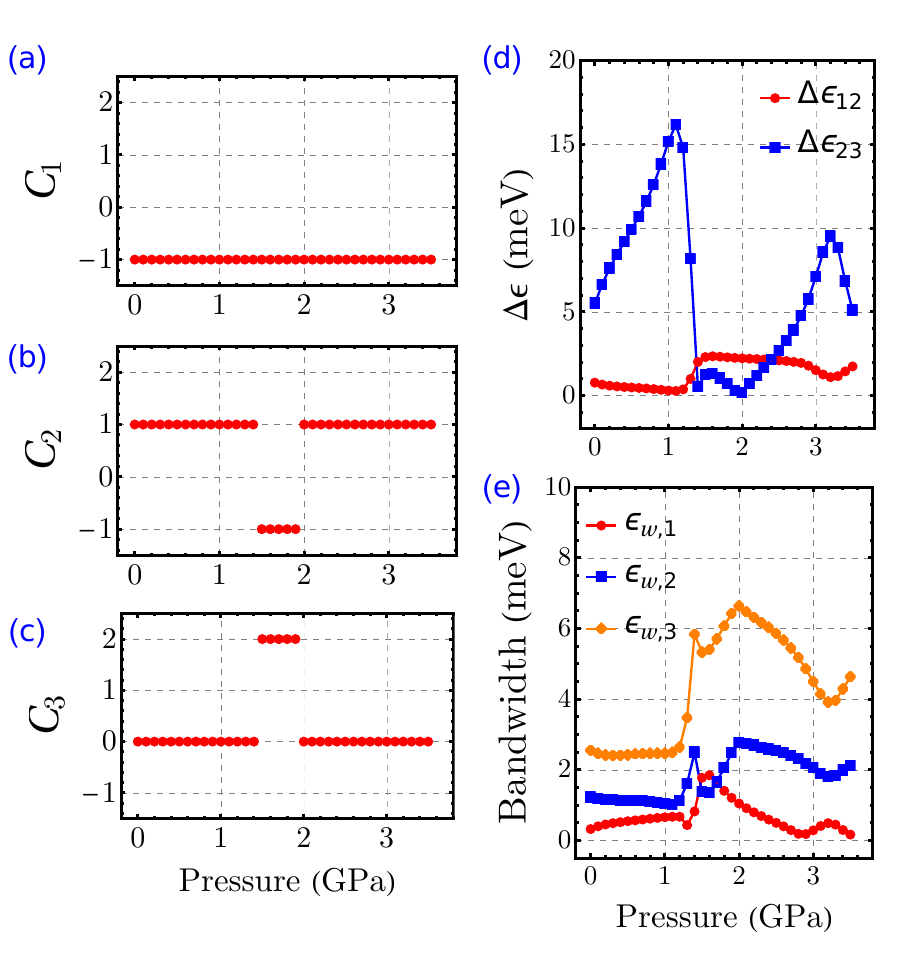}
	\caption{(a-c) Valley Chern numbers as a function of pressure, (d) the energy gap between the adjacent bands, and (e) the bandwidth of the top three bands for $\theta = 1.4^\circ$.}
	\label{fig:chern_1.4deg}
\end{figure}

As a first case, we consider a twist angle $\theta=1.4^\circ$. The bands and valley Chern numbers of the top three bands are shown in Fig. \ref{fig:chern_1.4deg}(a--c) while Fig. \ref{fig:chern_1.4deg}(d) shows the energy gaps between the adjacent bands $\epsilon_{ij}$ (calculated in the same way as in \ref{subsec:unpressurized-topology}). As in the unpressurized case, the first band has the valley Chern number $C_1$ of -1 and does not change along the pressure range. This behavior starkly contrasts the gap between the second and third bands. Here, bands start approaching each other from pressure $P = 1$ GPa, touching at a pressure of around $P=1.5$ GPa. Subsequently, valley Chern numbers change from $(C_2, C_3) = (1,0)$ to $(C_2, C_3) = (-1, 2)$. Another transition occurs when both bands touch again at a pressure $P \approx  2$ GPa. Here, valley Chern numbers change from $(C_2, C_3) = (-1, 2)$ to $(C_2, C_3) = (1, 0)$.

Fig. \ref{fig:chern_1.4deg}(e) shows the bandwidth of the three topmost bands. As can be seen, pressure affects the bandwidth of those bands. As an exciting sample, the width of the first band $\epsilon_{w,1}$ is less than 1 meV for pressure ranging from 0 -- 1.4 GPa. Although it increases to the maximum of 1.8 meV within the range of 1.4 -- 2 GPa, the bandwidth becomes narrow again after 2 GPa and reaches its minimum width of 0.12 meV at pressure $P = 3.5$ GPa as can be seen in Fig. \ref{fig:bands_1.4deg}. This type of band (with bandwidth $< 1$ meV) can be considered an ultra-flat band\cite{PhysRevB.102.075413}. Flat bands are interesting because they might indicate the relevancy of strong correlation effects for electrons occupying the corresponding band. The literature discusses various mechanisms that can lead to flat bands in twisted moire systems\cite{PhysRevLett.128.176404,Morales-Duran2023,PhysRevLett.122.106405}. Since the Hamiltonian discussed in \cite{Morales-Duran2023} is a close match to our Hamiltonian, one can expect that their description of a flat-band mechanism also applies to our work.

\subsubsection{\texorpdfstring{Twist Angle $\theta = 2.3^\circ$}{Twist Angle 2.3\textdegree}}

\begin{figure}[ht]
	\centering
	\includegraphics[width=\linewidth]{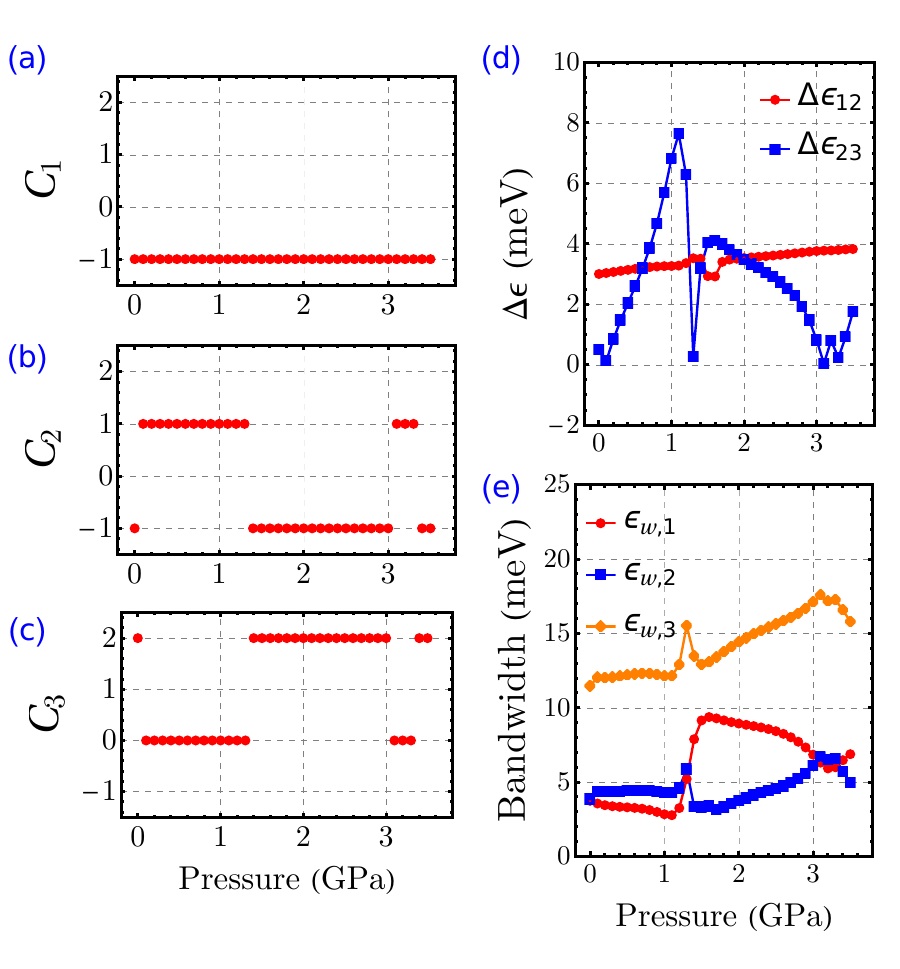}
	\caption{(a-c) Valley Chern numbers as a function of pressure, (d) the energy gap between the adjacent bands, and (e) the bandwidth of the top three bands for $\theta = 2.3^\circ$.}
	\label{fig:chern_2.3deg}
\end{figure}

\begin{figure}
	\centering
	\includegraphics[width=\linewidth]{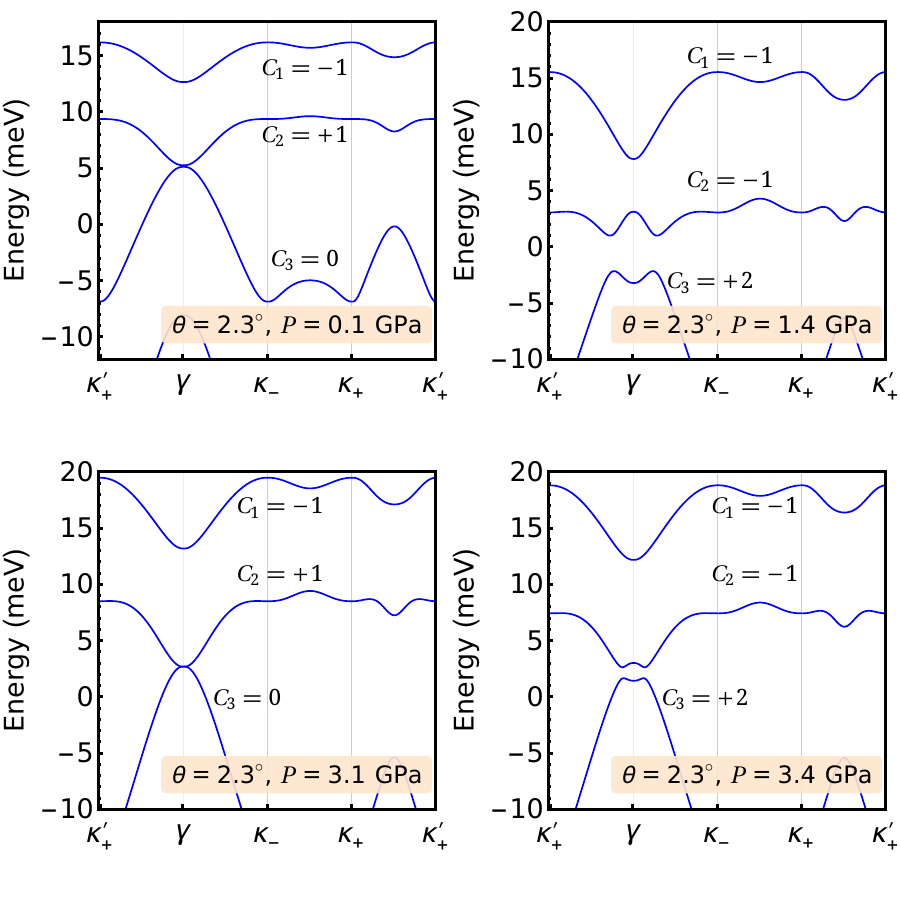}
	\caption{The band structure at various pressures for twist angle $\theta = 2.3^\circ$. In the figure, we have labeled the valley Chern number of the $n$-th band (counted from the top) as $C_n$.}
	\label{fig:bands_2.3deg}
\end{figure}

Next, we investigate the case of twist angle $\theta = 2.3^\circ$ which in the unpressurized case is the angle where the topological transition happens. As shown in \ref{fig:chern_2.3deg}(a-c), there are several transitions for the second and third bands. The first transition occurs immediately at $P = 0.1$ GPa after both bands touch at the $\boldsymbol{\gamma}$ point where the valley Chern numbers of the second and third bands change from $(C_2, C_3) = (-1,2)$ to $(C_2, C_3) = (1,0)$. Both bands cross again at $P \sim 1.3$ GPa, as shown in Fig. \ref{fig:chern_2.3deg}(d), and the band crossing is accompanied by a change in valley Chern numbers from $(C_2, C_3) = (1,0)$ to $(C_2, C_3) = (-1,2)$. Another band crossing occurs at the same point in $k$ space, i.e., $\gamma$, at pressure $P \sim 3.1$ GPa. Here, the band crossing is followed by a change of valley Chern numbers as $(C_2,C_3)=(-1,2) \to (1,0)$. Although both bands seem to cross in Fig. \ref{fig:bands_2.3deg}, the gap between the second and third bands is 0.03 meV. A final change of valley Chern numbers occurs after the crossing at $P \sim 3.3$ GPa as $(C_2, C_3) = (1, 0) \to (-1,2)$.

We expect that these pressure-induced changes in Chern number would manifest experimentally as pressure-induced changes in the Hall current for materials with appropriate band filling. 

We no longer observe any band with bandwidth $< 1$ meV, although the first and second bands still have bandwidth less than 10 meV. Their bandwidths are less than 5 meV (still considered flat by some literature standards \cite{Tao2022}), particularly in the region between 0 - 1.2 GPa. Apart from this range, the bandwidth of the second band also stays less than 5 meV in the range of 1.4 - 2.7 meV.

\section{Conclusion}
\label{sec:conclusion}

In summary, we started our work with a demonstration of the procedure to obtain the effective low-energy Hamiltonian for the twisted TMD based on a general symmetry-based approach \cite{Balents2019} and reproduced results that closely resemble those of Ref. \cite{wu2019topological}. We then use the energy data for pressurized untwisted TMDs from DFT computation to obtain an explicit form of the physical coupling parameters $V,\psi,w$ and $m^*$ as functions of pressure $P$ which are summarized in Eq. \eqref{eq:V_press}--\eqref{mf_press}. The model that we developed here is valid around the $\vect{K}$ region and can capture some physics of the system under pressure and further development is needed to obtain the contribution from other regions.

Using these parameters, we have also calculated the valley Chern numbers of the topmost three bands and observed multiple topological transitions as demonstrated by changes in valley Chern numbers for the second and third topmost bands. These changes occur at band crossings when the pressure reaches certain critical values. The occurrence of these transitions means that varying pressure can have a similar effect as varying the twist angle in the sense that both procedures give rise to non-trivial valley Chern numbers. Variations in pressures could be advantageous compared to changes in twist angle since an experimentalist could modify the material properties \emph{in situ} without the need to prepare different materials for each different twist angle to obtain the non-trivial topological insulating property

 An important outcome of our work is the derivation of the low-energy effective model that provides a numerically inexpensive (compared to brute force DFT methods) means for studying twisted bilayer materials under the influence of pressure. Typically, DFT computations of these materials would be prohibitively expensive because, at small twist angles, the unit cell contains many atoms, leading to a very high computational cost. Furthermore, the procedure of this work could provide a framework to study other types of TMD bilayer material. Our work makes it easy to study the interplay of external influences like electromagnetic fields and pressure in twisted TMD bilayers. It, therefore, opens the door for future studies in this direction. Another avenue for future work is the study of the effect of pressure on twisted TMD heterobilayers using similar methods. Our low-energy Hamiltonian could also potentially be used to study stretched or otherwise deformed TMDs. One could improve the model further by, for example, considering the effect of lattice relaxation for twisted configurations since we considered rigid twists only.

One of the merits of our work is that it clearly demonstrates that the physical properties of twisted TMD homobilayers are highly tunable through the use of a combination of pressure and twist angles.

\section{Acknowledgements}
M. V. and H. B. gratefully acknowledge the support provided by the Deanship of Research Oversight and Coordination (DROC) at King Fahd University of Petroleum \& Minerals (KFUPM) for funding this work through exploratory research grant No. ER221002. The Alfahd High-Performance Computing facility, provided by the College of Petroleum Engineering \& Geosciences at KFUPM, is acknowledged for its essential role in supporting the computational calculations.
\section{Data Availability}
Data that supports the findings of this article is available from the authors upon reasonable request.

\bibliographystyle{unsrt}
\bibliography{library}
\end{document}